\documentclass[]{tCPH2e}

\usepackage{color}
\usepackage{bbm}
\usepackage{graphicx,epsfig}

\newcommand{\tr}{{\rm tr}}
\newcommand{\de}{\delta}
\newcommand{\eps}{\epsilon}
\newcommand{\ini}{{\rm ini}}

\renewcommand{\d}{{\textrm{d}}}
\newcommand{\p}{{\partial}}
\newcommand{\upt}{^{(t)}}
\newcommand{\upz}{^{(0)}}
\newcommand{\upa}{^{(\tau)}}
\newcommand{\upo}{^{(1)}}
\newcommand{\upw}{^{(2)}}
\newcommand{\close}{\rm closed}

\newcommand{\Ren}	{S_{\alpha}}
\newcommand{\Reno}{S_{1}}
\newcommand{\Fa}{F_{\alpha}}

\def\<{\langle}
\def\>{\rangle}

\def\be{\begin{eqnarray}}
\def\ee{\end{eqnarray}}

\newcommand{\sai}[1]{{\color{black}#1}}
\newcommand{\add}[1]{\color{black}#1}
\newcommand{\sv}[1]{{\color{black}{#1}}}
\newcommand{\ja}[1]{{\color{black}{#1}}}
\newcommand{\jan}[1]{{\color{black}{#1}}}
\newcommand{\borrowedfig}[1]{{\color{black}{#1}}}

\begin{document}
%
%
%

%

\title{Quantum Thermodynamics}

\author{
Sai Vinjanampathy$^{a}$$^{\ast}$\thanks{$^\ast$Email: sai@quantumlah.org} \vspace{6pt}
and 
Janet Anders$^{b}$$^{\dag}$\thanks{$^\dag$Email:  janet@qipc.org}  \\ \vspace{6pt}  
$^{a}${\em{Centre for Quantum Technologies, National University of Singapore, 3 Science Drive 2, Singapore 117543.}}
$^{b}${\em{Department of Physics and Astronomy, University of Exeter, Stocker Road, EX4 4QL, United Kingdom.}}\\ \vspace{6pt} 
\received{~} 
}   
   
\maketitle

\begin{abstract}

Quantum thermodynamics is an emerging research field aiming to extend standard thermodynamics and non-equilibrium statistical physics to ensembles of sizes well below the thermodynamic limit, in non-equilibrium situations, and with the full inclusion of quantum effects. Fueled by experimental advances and the potential of future nanoscale applications this research effort is pursued by scientists with different backgrounds, including statistical physics, many-body theory, mesoscopic physics and quantum information theory, who bring various tools and methods to the field. A multitude of theoretical questions are being addressed ranging from issues of thermalisation of quantum systems and  various definitions of ``work'', to the efficiency and power of quantum engines. This overview provides a perspective on a selection of these current trends accessible to postgraduate students and researchers alike. 


\end{abstract}

\tableofcontents

\section{Introduction}
One of the biggest puzzles in quantum theory today is to show how the well-studied properties of a few  particles translate into a statistical theory from which new macroscopic quantum thermodynamic laws emerge. This challenge is addressed by the emerging field of \emph{quantum thermodynamics} which has grown rapidly over the last decade. \ja{It is fuelled by recent equilibration experiments \cite{Trotzky12} in cold atomic and other physical systems, the introduction of new numerical methods \cite{Scholl11}, and the discovery of fundamental theoretical relationships in non-equilibrium statistical physics and quantum information theory \cite{Jarzynski97, Crooks99, TASAKI,Sagawa-Ueda2010,PSW06,delRio,resource}.}
With ultrafast experimental control of quantum systems and engineering of small environments pushing the limits of conventional thermodynamics, the central goal of quantum thermodynamics is the extension of standard thermodynamics to include quantum effects and small ensemble sizes. 
Apart from the academic drive to clarify fundamental processes in nature, it is expected that industrial need for miniaturization of technologies to the nanoscale will benefit from understanding of quantum thermodynamic processes. Obtaining a detailed knowledge of how quantum fluctuations compete with thermal fluctuations is essential for us to be able to adapt existing technologies to operate at ever decreasing scales, and to uncover new technologies that may harness quantum thermodynamic features. 

Various perspectives have emerged in quantum thermodynamics, due to the interdisciplinary nature of the field, and each contributes different insights. For example, the study of thermalisation has been approached by quantum information theory from the standpoint of typicality and entanglement, and by many-body physics with a dynamical approach.  Likewise, the recent study of quantum thermal machines, originally approached from a quantum optics perspective \cite{scovil1959three,geusic1967quantum,scully2002quantum},  has since received significant input from many-body physics, fluctuation relations, and linear response approaches \cite{esposito2010quantum,mazza2014thermoelectric}. These designs further contrast with studies on thermal machines based on quantum information theoretic approaches \cite{gemmer2003distribution,GMM09,schulman1999molecular,linden2010small,PhysRevE.86.040106,jacobs2014quantum}. The difference in perspectives on the same topics has also meant that there are ideas within quantum thermodynamics where consensus is yet to be established.

This article is aimed at non-expert readers and specialists in one subject who seek a brief overview of  quantum thermodynamics. The scope is to give an introduction to some of the different perspectives on current topics in a single paper, and guide the reader to a selection of useful references. Given the rapid progress in the field there are many aspects of quantum thermodynamics that are not covered in this overview. Some topics have been intensely studied for several years and dedicated reviews are available, for example, in classical non-equilibrium thermodynamics \cite{de2013non}, fluctuation relations \cite{campisi2011colloquium}, non-asymptotic quantum information theory \cite{tomamichel2012framework},  quantum engines \cite{kosloff2014quantum}, \sv{equilibration and thermalisation \cite{dunjko2012thermalisation,gogolin2015equilibration}}, and a recent quantum thermodynamics review focussing on quantum information theory techniques \cite{Goold15}. \ja{Other reviews of interest discuss Maxwell's demon and the physics of forgetting \cite{VP01, Maruyama09,Ford16} and thermodynamic aspects of information \cite{PHS}.} We encourage researchers to take on board the insights gained from different approaches and attempt to fit together pieces of the puzzle to create an overall united framework of quantum thermodynamics. 

Section \ref{sec:qt} discusses the standard laws of thermodynamics and introduces the link with information processing tasks and Section \ref{sec:cqnesp} gives a brief overview of fluctuation relations in the classical and quantum regime. Section \ref{sec:qdft} then introduces quantum dynamical maps and discusses implications for the foundations of thermodynamics. Properties of maps form the backbone of the thermodynamic resource theory approach to single shot thermodynamics discussed in Section \ref{sec:sst}. Section \ref{sec:Tmachines} discusses the operation of thermal machines in the quantum regime. Finally, in Section \ref{sec:discussion} the current state of the field is summarised and open questions are identified.

\section{Information and Thermodynamics} \label{sec:qt}

This section defines averages of heat and work, introduces the first and second law of thermodynamics and discusses examples of the link between thermodynamics and information processing tasks, such as erasure. 

\subsection{The first and second law of thermodynamics} \label{sub:1st+2ndlaw}

Thermodynamics is concerned with energy and changes of energy that are distinguished as heat and work. For a quantum system in state $\rho$ and with Hamiltonian $H$ at a given time the system's internal, or \emph{average}, energy is identified with the expectation value $U(\rho) = \tr[\rho \, H]$. When a system changes in time, i.e. the pair of state and Hamiltonian \cite{AG13} vary in time $(\rho\upt, H\upt)$ with $t \in [0, \tau]$, the resulting average energy change
\be 
	\Delta U = \tr[\rho\upa \, H\upa] - \tr[\rho\upz \, H\upz]
\ee 
is made up of two types of energy transfer - \ja{work and heat}. Their intuitive meaning is that of two types of energetic resources, one fully controllable and useful, the other uncontrolled and wasteful \cite{PW,LENARD,ALICKI,KR,TASAKI,KIEW,ALLA,ESPOSITO,HENRICH, AG13}. Since the time-variation of $H$ is controlled by an experimenter the energy change associated with this time-variation is identified as \emph{work}. The uncontrolled energy change associated with the reconfiguration of the system state in response to Hamiltonian changes and the system's coupling with the environment is identified as \emph{heat}. The formal definitions of \emph{average heat absorbed by the system} and \emph{average work done on the system} are then
\be \label{eq:Q} \label{eq:Q+W} 
	\<Q\>  := \int_{0}^{\tau}  \tr[ \dot \rho\upt \,  H\upt] \, \d t  \quad \mbox{and} \quad 
	\<W\> := \int_{0}^{\tau}  \tr[ \rho\upt \,  \dot {H}\upt] \, \d t. 
\ee
Here the brackets $\< \cdot \>$ indicate the ensemble average that is assumed in the above definition when the trace is performed. Work is extracted from the system when $\<W_{\rm ext}\> := - \< W\> > 0$, while heat is dissipated to the environment when $\<Q_{\rm dis}\> := - \< Q\> > 0$. 

The \emph{first law of thermodynamics}  states that the sum of average heat and work done on the system just makes up its average energy change,
\be \label{eq:firstlaw}
	\< Q \> + \< W\> =  \int_{0}^{\tau}  {\d \over \d t} \tr[ \rho\upt \,  H\upt]  \, \d t 
	=  \tr[ \rho\upa \,  H\upa] -  \tr[ \rho\upz \,  H\upz]
	= \Delta U.
\ee 
It is important to note that while the internal energy change only depends on the initial and final states and Hamiltonians of the evolution, heat and work are process dependent, i.e. it matters \emph{how} the system evolved in time from $(\rho\upz, H\upz)$ to  $(\rho\upa, H\upa)$. Therefore heat and work for an infinitesimal process will be denoted by $\< \de Q\>$ and $\<\de W\>$ where the symbol $\de$ indicates that heat and work are (in general) \emph{not} full differentials and do not correspond to observables \cite{Talkner2007}, in contrast to the average energy with differential $\d U$. 

Choosing to \emph{split} the energy change into two types of energy transfer is crucial to allow the formulation of the second law of thermodynamics. A fundamental law of physics, it sets limits on the work extraction of heat engines and establishes the notion of \emph{irreversibility} in physics. 
Clausius observed in 1865 that a new state function - the \emph{thermodynamic entropy} $S_{\rm th}$ of a system - is helpful to study the heat flow to the system when it interacts with baths at varying temperatures $T$ \cite{FordBook}. The thermodynamic entropy is defined through its change in a \emph{reversible thermodynamic} process, 
\be \label{eq:Sthdef}
	\Delta S_{\rm th} :=\int_{rev} {\< \de Q \> \over T},
\ee
where $\< \de Q \>$ is the heat absorbed by the system along the process and $T$ is the temperature at which the heat is being exchanged between the system and the bath.  Further observing that any cyclic process obeys $\oint {\< \de Q \>  \over T} \le 0$ with equality for reversible processes Clausius formulated a version of the \emph{second law of thermodynamics} for all thermodynamic processes, today known as the Clausius-inequality:
\be \label{eq:Clausius}
	\int {\< \de Q \> \over T} \le \Delta S_{\rm th}, \quad \mbox{becoming} \quad \<  Q \>  \le T \, \Delta S_{\rm th} \quad \mbox{for $T$ = const}.
\ee
It states that the change in a system's entropy must be equal or larger than the average heat absorbed by the system during a process \ja{divided by the temperature at which the heat is exchanged}. In this form Clausius' inequality establishes the existence of an upper bound to the heat absorbed by the system and its validity is generally assumed to extend to the quantum regime \cite{PW,LENARD,ALICKI,KR,TASAKI,KIEW,ALLA,ESPOSITO,HENRICH, GMM09,AG13}. Equivalently, by defining the \emph{free energy} of a system with Hamiltonian $H$ and in contact with a heat bath at temperature $T$ as the state function
\be \label{eq:FE}
  	F(\rho) := U(\rho) - T S_{\rm th} (\rho),
\ee 
Clausius' inequality becomes a statement of the upper bound on the work that can be extracted in a thermodynamic process, 
\be \label{eq:Worklaw}
	 \< W_{\rm ext} \> = - \< W \> = - \Delta U +  \<  Q \> \le - \Delta U + T \Delta S_{\rm th} = - \Delta F.
\ee
While the actual heat absorbed/work extracted will depend on the specifics of the process there exist optimal, thermodynamically reversible, processes that saturate the equality, see Eq.~(\ref{eq:Sthdef}). \ja{However, modifications of the second law, and thus the optimal work that can be extracted, arise when the control of the working system is restricted to physically realistic, local scenarios \cite{WRE14}. }

For equilibrium states $\rho_{\rm th} = e^{-\beta H}/\tr[e^{-\beta H}]$ for Hamiltonian $H$ and at inverse temperatures $\beta = {1 \over k_B \, T}$ the thermodynamic entropy $S_{\rm th}$ equals the information theory entropy,  $S$,  \ja{times the Boltzmann constant $k_B$, i.e.  ${S_{\rm th} (\rho_{\rm th})  } = k_B \, S (\rho_{\rm th})$. The information theory entropy, known as the Shannon or von Neumann entropy, for a general state $\rho$ is defined as}
\be \label{eq:vNE}
	S (\rho) : = - \tr [\rho \, \ln \rho].
\ee
\jan{Many researchers in quantum thermodynamics assume that the thermodynamic entropy is naturally extended to non-equilibrium states by the information theoretic entropy. For example, this assumption is made when using the von Neumann entropy in connection with the second law and the analysis of thermal processes, and in the calculation of efficiencies of quantum thermal machines. Evidence that this extension is appropriate has been provided via many routes including Landauer's original work, see \cite{VP01} for an introduction. The suitability of this extension, and its limitations, remain however debated issues \cite{entropy1, entropy2}.  For the remainder of this article we will assume that the von Neumann entropy $S$ is the natural extension of the thermodynamic entropy $S_{\rm th}$. }

\subsection{Maxwell's demon} \label{sub:demon}

\begin{figure}[t]
\centering
 \includegraphics[width=0.5\textwidth]{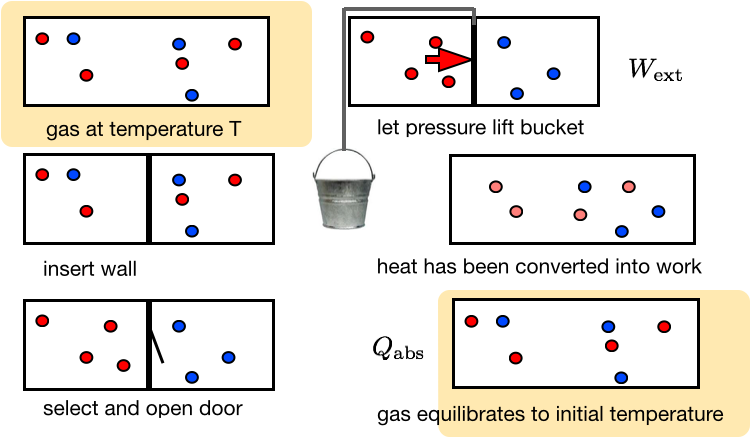} 
	\caption{\label{fig:demon} A gas in a box starts with a thermal distribution at temperature $T$. Maxwell's demon inserts a wall and selects faster particles to pass through a door in the wall to the left, while he lets slower particles pass to the right, until the gas is separated into two boxes that are not in equilibrium. The demon attaches a bucket (or another suitable work storage system) and allows the wall to move under the pressure of the gases. Some gas energy is extracted as work, $W_{\rm ext}$, raising the bucket. Finally, the gas is brought in contact with the environment at temperature $T$, equilibrating back to its initial state ($\Delta U = 0$) while absorbing heat $Q_{\rm abs} = W_{\rm ext}$. A comprehensive review of Maxwell's demon is \cite{Maruyama09}. }
\end{figure}

Maxwell's demon is a creature that is able to observe the motion of individual particles and use this information (by employing feedback protocols discussed in section \ref{sub:feedback}) to convert heat into work in a cyclic process \ja{using only a single heat bath at temperature $T$.} By separating slower, ``colder'' gas particles in a container from faster, ``hotter'' particles, and then allowing the hotter gas to expand while pushing a piston, see Fig.~\ref{fig:demon}, the demon can extract work while returning to the initial mixed gas. For a single particle gas, the demon's extracted work in a cyclic process is
\be \label{eq:demon}
	\< W^{\rm demon}_{\rm ext} \>  = k_B T \, \ln 2.
\ee
Maxwell realised in 1867 that such a demon would appear to break the second law of thermodynamics, \ja{see Eq.~(\ref{eq:Clausius}),} as it converts heat completely into work in a cyclic process resulting in a positive \emph{extracted work}, $\<W^{\rm demon}_{\rm ext}\> =  \<Q\> - \Delta U \not \le T \Delta S_{\rm th} - \Delta U = 0$. The crux of this paradoxical situation is that the demon acquires \emph{information} about individual gas particles and uses this information to convert heat into useful work.  One way of resolving the paradox was presented by Bennett \cite{Bennett82} and invokes Landauer's erasure principle \cite{Landauer61}, as described in the next section \ja{and presented highly accessibly in \cite{VP01}. Other approaches consider the cost the demon has to pay upfront when identifying whether the particle is slower or faster \cite{Ford16}. Experimentally,  the thermodynamic phenomenon of Maxwell's demon remains highly relevant - for instance, cooling a gas can be achieved by mimicking the demonic action \cite{Raizen11}. An example of an experimental test of  Maxwell's demon is discussed in section \ref{sub:clexp}.}

\subsection{Landauer's erasure principle}  \label{sub:erasure}

In a seminal paper \cite{Landauer61} Landauer investigated the thermodynamic cost of information processing tasks. He concluded that the erasure of one bit of information requires a minimum dissipation of heat,
\be \label{eq:Landauer}
	\< Q^{\min}_{\rm dis} \>  = k_B \, T \, \ln 2,
\ee
that the erased system dissipates to a surrounding environment, cf. Fig.~\ref{fig:demon}, in equilibrium at temperature $T$. The erasure of one bit, \ja{ or ``reset'', here refers to the change of a system being in one of two states with equal probability, ${\mathbbm{1}\over2}$ (i.e. 1 bit), to a definite known state, $|0 \>$ (i.e. 0 bits). (We note that, contrary to everyday language, the technical term ``information'' here refers to uncertainty rather than certainty. Erasure of information thus implies increase of certainty.)} Landauer's principle has been tested experimentally only very recently, see section \ref{sub:clexp}. 

The energetic cost of the dissipation is balanced by a minimum work that must be done on the system, $\<W^{\rm min}\> = \Delta U - \< Q\> =  \< Q^{\min}_{\rm dis}\> $, to achieve the erasure at constant average energy, $\Delta U =0$. Bennett argued \cite{Bennett82} that Maxwell's demonic paradox can thus be resolved by taking into account that the demon's memory holds bits of information that need to be erased to completely close the thermodynamic cycle and return to the initial conditions. The work that the demon extracted in the first step, see previous section, then has to be spent to erase the information acquired, with no net gain of work and in agreement with the second law.

While Landauer's  principle was originally formulated for the erasure of one bit of classical information, it is straightforwardly extended to the erasure of a general mixed quantum state, $\rho$, which is transferred to the blank state $|0 \>$. The minimum required heat dissipation is then $\<Q^{\min}_{\rm dis} \> = k_B \, T \, S(\rho)$ following from the second law, Eq.~(\ref{eq:Clausius}). \ja{A recent analysis of Landauer's principle \cite{RW14} uses a framework of \emph{thermal operations}, cf. section \ref{sec:sst}, to obtain corrections to Landauer's bound when the effective size of the thermal bath is finite. They show that the dissipated heat is in general above Landauer's bound and only converges to it for infinite-sized reservoirs. Probabilistic erasure is considered in \cite{MMO16} and the tradeoff between the probability of erasure and minimal heat dissipation is demonstrated. In the simplest case the erasure protocol achieving the Landauer bound will require an idealised quasi-static process which would take infinitely long to implement. The time of erasure is investigated in \cite{BGDV14} and it is shown that Landauer's limit is achievable in finite time when allowing exponentially suppressed erasure errors. }

\subsection{Erasure with quantum side information} \label{sub:sideinfo}

The heat dissipation during erasure is non-trivial when the initial state of the system is a mixed state, $\rho_S$. In the \emph{quantum regime} mixed states can always be seen as reduced states of global states, $\rho_{SM}$, of the system $S$ and a memory $M$, with $\rho_S = \tr_M[\rho_{SM}]$. A groundbreaking paper  \cite{delRio} takes this insight seriously and sets up an erasure scenario where an observer can operate on both, the system and the memory. During the global process the system's local state is erased, $\rho_S \mapsto |0\>$, while the memory's local state, $\rho_M = \tr_S[\rho_{SM}]$, is not altered. In other words, this global process is locally indistinguishable from the erasure considered in the previous section. However, contrary to the previous case, erasure with ``side-information'' \ja{, i.e. using correlations of the memory with the system,} can be achieved while \emph{extracting} a maximum amount of work
\be \label{eq:erasureside}
	\< W^{\max}_{\rm ext} \>  = - k_B \, T \, \jan{S(S|M)_{\rho_{SM}}}.
\ee
\jan{Here $S(S|M)_{\rho_{SM}}$ is the conditional von Neumann entropy between the system and memory, $S(S|M)_{\rho_{SM}} = S(\rho_{SM}) - S(\rho_M)$.} Crucially, the \emph{conditional entropy can be negative} for some quantum correlated states (a subset of the set of entangled states) thus giving a positive extractable work. This result contrasts strongly with Landauer's principle valid for both classical and quantum states when no side-information is available. \jan{The possibility to extract work during erasure is a purely quantum feature, that relies on accessing the side-information \cite{delRio}. I.e. to practically obtain positive work requires knowledge of and access to an initial entangled state of the system and the memory, and the implementation of a carefully controlled process on the degrees of freedom of both parties. The entanglement between system and memory will be destroyed in the process and can be seen as ``fuel'' from which work is extracted. }

\medskip

\subsection{Work from correlations} \label{sub:wfcorr}

\jan{The thermodynamic work and heat associated with creating or destroying (quantum) \emph{correlations} has been studied intensely, e.g. \cite{Opp02, Zur03,DL09,DRRV11,JJR12,FWU13,PL15,Huber15}, for a variety of settings, including  unitary and non-unitary processes. For example, the thermodynamic efficiency of an engine operating on pairs of correlated atoms can be quantified in terms of quantum discord and it was shown to exceed the classical efficiency value \cite{DL09}. In \cite{HSAL11} the minimal  heat dissipation for coupling a harmonic oscillator that starts initially in local thermal equilibrium, and ends up correlated with a bath of harmonic oscillators in a \emph{global} thermal equilibrium state, is determined and it was shown that this heat contribution resolves a previously reported second law violation. 

Thermodynamic aspects of creating correlations are also studied in \cite{Huber15} where the minimum work cost is established for \emph{unitarily} evolving an initial, locally thermal, state of $N$ systems to a global correlated state. A maximum temperature is derived at which entanglement can still be created, along with the minimal associated energy cost. In turn, when wanting to extract work from many-body states that are initially globally correlated, while locally appearing thermal, the maximum extractable work under global \emph{unitary} evolutions is discussed in \cite{PL15} for initial entangled and separable states, and those diagonal in the energy basis. This contrasts with the non-unitary process of erasure with side-information, discussed in subsection \ref{sub:sideinfo}.
}

\ja{
\subsection{Work from coherences} \label{sub:cohW}

\jan{While erasure with side-information discussed in subsection \ref{sub:sideinfo} and the results in subsection \ref{sub:wfcorr} illustrate} the quantum thermodynamic aspect of correlations, a second quantum thermodynamic feature arises due to the presence of coherences.  A recent paper \cite{KA15} identifies projection processes as a route to analyse the thermodynamic role of coherences. Projection processes map an initial state, $\rho$, which has coherences in a particular basis of interest, $\{ \Pi_k \}_k$ \ja{with $k=1, 2, 3, ...$,} to a state in which these coherences are removed, i.e.  $\rho \mapsto \eta := \sum_k \Pi_k \, \rho \, \Pi_k$. These processes can be interpreted as \emph{unselective measurements} of an observable with eigenbasis  $\{ \Pi_k \}_k$, a measurement in which not the individual measurement outcomes are recorded but only the statistics of the outcomes is known. Just like in the case of Landauer's erasure map, the state transfer achieved by the new mathematical map can be implemented in various physical ways. Different physical implementations of the same map will have different work and heat contributions. It was shown that there exists a physical protocol to implement the projection process such that a non-trivial average work can be extracted from the initial state's coherences. For the example that the basis, $\{ \Pi_k \}_k$, is the energy eigenbasis of a Hamiltonian $H = \sum_k E_k \, \Pi_k$ with eigenenergies $E_k$ the \emph{maximum amount of average work} that can be extracted in a projection process is
\be \label{eq:measure}
	\< W^{\max}_{\rm ext} \>  = k_B \, T \, (S(\eta) - S(\rho)) \ge 0.
\ee
Importantly $S(\eta)$ is larger than $S(\rho)$ \emph{if and only if} the state $\rho$ had coherences with respect to the energy eigenbasis, while the entropy does not change under projection processes for classical states. Thus the extracted work here is due to the  \emph{quantum coherences} in the initial state, contrasting with the work gained from quantum correlations discussed in subsections \ref{sub:sideinfo} and \ref{sub:wfcorr}. 
We note that ``decohering'' a state $\rho$ is a physical implementation of the same state transfer, $\rho \mapsto \eta$, during which coherences are washed out by the environment in an uncontrolled way and no work is extracted. This is a suboptimal process - to achieve the maximum work and optimal implementation needs to be realised which requires a carefully controlled protocol of interacting the system with heat and work sources \cite{KA15}. It is worth noting that in contrast to Maxwell's demon whose work extraction is based on the knowledge of ``microstates'' and appears to violate the second law when information erasure is not considered, gaining work from coherences is in accordance with the second law. Once the projection process is completed the final state has lost its coherence, i.e. here the coherences have been used as ``fuel'' to extract work. 
}


\section{Classical and Quantum Non-Equilibrium Statistical Physics} \label{sec:cqnesp}

In this section the statistical physics approach is outlined that  rests on the definition of fluctuating heat and work from which it derives the ensemble quantities of average heat and average work discussed in section \ref{sec:qt}. Fluctuation theorems and experiments are first described in the classical regime before they are extended to the quantum regime.

\subsection{Definitions of classical fluctuating work and heat} \label{subsec:Fluc}

In classical statistical physics a single particle is assigned a point, $x = (q, p)$, in phase space while an ensemble of particles is described with a probability density function, $P(x)$, in phase space.  The Hamiltonian of the particle is denoted as $H (x, \lambda)$ where $x$ is the phase space point of the particle for which the energy is evaluated and $\lambda$ is an externally controlled force parameter that can change in time. For example, a harmonic oscillator Hamiltonian $H (x, \lambda) = {p^2 \over 2m} + {m \lambda^2 q^2 \over 2}$ can become time-dependent through a \emph{protocol} according to which the frequency, $\lambda$, of the potential is varied in time, $\lambda(t)$. This particular example will be  discussed further in section \ref{sec:Tmachines} on quantum thermal machines. 

\begin{figure}[t]
\centering
	\includegraphics[width=0.35\textwidth]{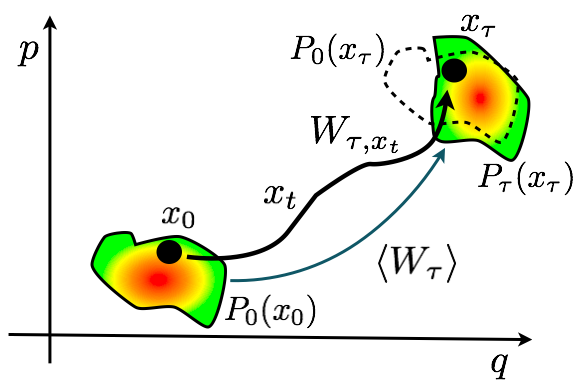} 
	\caption{\label{fig:phasespace} Phase space spanned by position and momentum coordinates, $ x= (q, p)$. A single trajectory, $x_t$, starts at phase space point $x_0$ and evolves in time $t$ to a final phase space point $x_{\tau}$. This single trajectory has an associated fluctuating work $W_{\tau, x_t}$. In general, an ensemble of initial phase space points described by an initial probability density functions, $P_0 (x_0)$, evolves to some final probability density function $P_{\tau} (x_{\tau})$. The ensemble of trajectories has an associated average work $\< W_{\tau}\>$. \ja{For closed dynamics the final probability distribution $P_0 (x_{\tau})$ (dashed line) has the same functional dependence as the initial probability distribution $P_0 (x_0)$ just evaluated at the final phase space point $x_{\tau}$. }
	}
\end{figure}

Each particle's state will evolve due to externally applied forces and due to the interaction with its environment resulting in a trajectory $x_t = (q_t, p_t)$ in phase space, see Fig.~\ref{fig:phasespace}. The \emph{fluctuating work} (done on the system) and the \emph{fluctuating heat} (absorbed by the system) for a single trajectory $x_t$ followed by a particle in the time window $[0, \tau]$ \ja{are} defined, analogously to (\ref{eq:Q+W}), as the energy change due to the externally controlled force parameter and the response of the system's state,
\be \label{eq:work+heat}
	W_{\tau, x_t} := \int_0^{\tau}  {\p H (x_t, \lambda(t)) \over \p \lambda(t)} \, \dot \lambda(t) \, \d t 
	\quad \mbox{and} \quad
	Q_{\tau, x_t}  := \int_0^{\tau} {\p H (x_t, \lambda(t)) \over \p x_t} \, \dot x_t \, \d t . 
\ee
These are stochastic variables that depend on the particle's trajectory $x_t$ in phase space. Together they give the energy change of the system, cf. (\ref{eq:firstlaw}), along the trajectory 
\be \label{eq:fluc1stlaw}
	W_{\tau, x_t} + Q_{\tau, x_t} = H (x_{\tau}, \lambda({\tau})) - H (x_0, \lambda(0)).
\ee
For a \emph{closed system} one has
\be \label{eq:zeroQ}
 	Q^{\close}_{\tau, x_t} 
	&=& \int_0^{\tau}  \left( {\p H (x_t, \lambda(t)) \over \p q_t} \, \dot q_t  +  {\p H (x_t, \lambda(t)) \over \p p_t} \, \dot p_t \right)  \d t  
	= \int_0^{\tau} \left( - \dot p_t  \dot q_t  +  \dot q_t  \dot p_t \right)  \d t  = 0,   \nonumber \\
	W^{\close}_{\tau, x_t} 
	&=& 	H (x_{\tau}, \lambda({\tau})) - H (x_0, \lambda(0)),
\ee
since the \ja{(Hamiltonian)} equations of motion, $ {\p H \over \p q} = - \dot p$ and $ {\p H \over \p p} = \dot q$, apply to closed systems. This means that a closed system has no heat exchange (a tautological statement)  and that the change of the system's energy during the protocol is identified entirely with work for each  single trajectory.

The \emph{average work} of the system refers to repeating the same experiment many times each time preparing the same initial system state, \ja{bringing the system in contact with the same kind of environment} and applying the same protocol. The average work is mathematically obtained as the integral over all trajectories, $\{x_t\}$, explored by the system, $ \< W_{\tau} \> : = \int P_t(x_t) \, W_{\tau, x_t}  \, \ja{\d {\cal D}}  (x_t)$, where $P_t(x_t)$ is the probability density of a trajectory $x_t$ and $\ja{\d {\cal D}} (x_t)$ the phase space integral over all trajectories \cite{Sagawa-Ueda2010}. 
\ja{For a \emph{closed system} the trajectories are fully determined by their initial phase space point alone: each trajectory starting with $x_0$ evolves \emph{deterministically} to $x_{\tau}$ at time $\tau$. Thus the probability of the trajectory is given by the initial probability $P_0(x_0)$ of starting in phase space point $x_0$, $\d {\cal D} (x_t) \to \d x_0$, and the Jacobian determinant is ${\d x_0 \over \d x_{\tau}} =1$.}  Combining with Eqs.~(\ref{eq:work+heat}), (\ref{eq:fluc1stlaw}) and (\ref{eq:zeroQ}), one obtains 
\be 	\label{eq:closedW}
	\< W^{\close}_{\tau} \>
	&=& \int P_0(x_0) \, \left( H(x_{\tau} (x_0), \lambda(\tau)) - H(x_0, \lambda(0)) \right) \, \d x_0, \nonumber \\
	&=& \int P_0 (x_\tau) \, H(x_{\tau}, \lambda(\tau)) \,  \left|  {\d x_0 \over \d x_{\tau} }\right| \, \d x_{\tau} 	
	- \int P_0(x_0) \, H(x_0, \lambda(0))  \, \d x_0, \\
	&=& U_{\tau}  - U_0, \nonumber
\ee
i.e. the average work for a closed process is just the difference of the average energies. \ja{Because the evolution is closed the distribution in phase space moves but keeps its volume, cf. Fig.~\ref{fig:phasespace}; which is known as Liouville's theorem. The final probability distribution, $P_0 (x_{\tau})$, has indeed the same functional form as the initial probability distribution, $P_0(x_0)$, just applied to the final trajectory points $x_{\tau}$. I.e. the probability of finding $x_{\tau}$ at the end is exactly the same as the probability of finding $x_0$ initially. In contrast, the probability distribution would change in time in open dynamics due to the system's non-deterministic interaction with the environment. In (\ref{eq:closedW})  the initial average energy of the system is defined as $U_{0} = \int P_{0} (x) \, H(x, \lambda(0))  \, \d x$ and similarly $U_{\tau}$ is the average energy at time $\tau$. }


\medskip

In statistical physics experiments the average work is often established by measuring the fluctuating work $W_j$ for a trajectory observed in a particular run $j$ of the experiment, repeating the experiment $N$ times, and taking the arithmetic mean of the results. In the limit $N \to \infty$ this is equivalent to constructing (from a histogram of the outcomes $W_j$) the probability density distribution for the work, $P(W)$, and then averaging with this function to obtain the average work
\be \label{eq:trajW2}
 	\< W_{\tau} \>  = \lim_{N \to \infty} {1 \over N} \sum_{j=1}^N W_j =  \int P (W) \, W \, \d W.
\ee

\subsection{Classical fluctuation relations} \label{sub:CFR}

A central recent breakthrough in classical statistical mechanics is the extension of the second law of thermodynamics, an inequality, to equalities valid for large classes of non-equilibrium processes. \emph{Detailed fluctuation relations} show that the probability densities of stochastically fluctuating quantities for a non-equilibrium process, such as entropy, work and heat, are linked to equilibrium properties and corresponding quantities for the time-reversed process \cite{ECM93, Crooks99, KPB07}. By integrating over the probability densities one obtains \emph{integral fluctuation relations}, such as Jarzynski's work relation \cite{Jarzynski97}. 

An important detailed fluctuation relation is \emph{Crooks relation} for a system in contact with a bath at inverse temperature $\beta = 1/(k_B T)$ for which detailed balanced is valid \cite{Crooks99}. The relation links the work distribution $P^F(W)$ associated with a \emph{forward} protocol changing the force parameter $\lambda(0) \to \lambda(\tau)$, to the work distribution $P^B(- W)$ associated with the time-reversed \emph{backwards} protocol where $\lambda(\tau) \to \lambda(0)$,
\be \label{eq:Crooks}
	P^F(W) = P^B(- W) \, e^{\beta (W-  \Delta F)}.
\ee
\ja{Here the forward and backward protocol each start with an initial distribution that is thermal at inverse temperature $\beta$ for the respective values of the force parameter. }
The free energy difference $\Delta F = F\upa - F\upz $ refers to the two thermal distributions with respect to the final, $H (x, \lambda(\tau))$, and initial, $H (x, \lambda(0))$, Hamiltonian in the forward protocol at inverse temperature $\beta$. Here the free energies, defined in (\ref{eq:FE}), can be expressed as $F\upz = - {1 \over \beta} \ln Z\upz$ at time 0 and similarly at time $\tau$ with classical partition functions $Z\upz := \int e^{- \beta H(x, \lambda(0))}  \, \d x$ and similarly $Z\upa$. Rearranging and integrating over $\d W$ on both sides results in a well-known integral fluctuation relation, the \emph{Jarzynski equality} \cite{Jarzynski97},
\be 
	\langle e^{- \beta W} \rangle = \int P^F (W) \, e^{- \beta W} \, \d W= e^{ - \beta \Delta F}.
\ee  

Pre-dating Crooks relation, Jarzynski proved his equality \cite{Jarzynski97} by considering a \emph{closed system} that starts with a \emph{thermal distribution}, $P_0(x_0) = {e^{- \beta H(x_0, \lambda(0))} \over Z\upz}$, for a given Hamiltonian $H(x, \lambda(0))$ at inverse temperature $\beta$. The Hamiltonian is externally modified through its force parameter $\lambda$ which drives the system out of equilibrium and into evolution according to Hamiltonian dynamics in a time-interval $[0, \tau]$. 
\ja{Such experiments can be realised, e.g. with colloidal particles see subsection \ref{sub:clexp}, where the experiment is repeated many times, each time implementing the same force protocol. Averages can then be calculated over many trajectories each starting from an initial phase space point that was sampled from an initial thermal distribution. The average exponentiated work done on the system is then obtained similarly to 
 (\ref{eq:closedW})} and one obtains
\be 	\nonumber 
	\< e^{- \beta W^{\close}_{\tau}} \>
	&=& \int P_0(x_0) \,  e^{- \beta W^{\close}_{\tau, x_\tau}} \,  \d x_0 
	= \int {e^{- \beta H(x_0, \lambda(0))} \over Z\upz} \,  e^{- \beta (H(x_\tau, \lambda(\tau)) - H(x_0, \lambda(0)) )} \,  \d x_0, \\
	&=&  {1 \over Z\upz} \, \int e^{- \beta H(x_\tau, \lambda(\tau))}  \,  \left|  {\d x_0 \over \d x_{\tau} }\right|\, \d x_{\tau} 
	= {Z\upa \over Z\upz} = e^{ - \beta \Delta F}. \label{eq:Jarzynski}
\ee
The beauty of this equality is that for all closed but arbitrarily strongly non-equilibrium processes the \ja{average exponentiated work} is determined entirely by equilibrium parameters contained in $\Delta F$. 
As Jarzynski showed the equality can also be generalised to open systems \cite{Jarzynski04}.
By applying Jensen's inequality $\langle e^{- \beta W} \rangle \ge e^{- \beta \langle W \rangle}$, the Jarzynski equality turns into the standard second law of thermodynamics, cf. Eq.~(\ref{eq:Worklaw}),
\be
	\< W \> \ge \Delta F.
\ee
Thus Jarzynski's relation strengthens the second law by including all moments of the non-equilibrium work resulting in an \emph{equality} from which the inequality follows for the first moment. 

Fluctuation theorems have been measured, for example, for a defect centre in diamond \cite{SSTWS05}, for a torsion pendulum \cite{DJGPC06}, and in an electronic system \cite{Saira12}. They have also been used to infer, from the measurable work in non-equilibrium pulling experiments, the desired equilibrium free-energy surface of bio-molecules \cite{Liphardt02, Collin05}, which is not directly measurable otherwise. 

\subsection{Fluctuation relations with feedback} \label{sub:feedback}

It is interesting to see how Maxwell demon's feedback process discussed in subsection \ref{sub:demon} can be captured in a generalised Jarzynski equation \cite{Sagawa-Ueda2010}. \ja{Again one assumes that the system undergoes closed dynamics with a time-dependent Hamiltonian, however, now the force parameter in the Hamiltonian is changed according to a protocol that \emph{depends} on the phase space point the system is found in. For example, for the two choices of a particle being in the left box or the right box, see Fig.~\ref{fig:Maxwell}, the initial Hamiltonian $H(x, \lambda(0))$ will be changed to either $H(x, \lambda_1(\tau))$ for the particle in the left box or $H(x, \lambda_2(\tau))$ for the particle in the right box. Calculating the average exponentiated work for this situation, see Eq.~(\ref{eq:Jarzynski}), one now notices that the integration over trajectories includes the two different evolutions, driven by one of the two Hamiltonians. 
The free energy difference, $\Delta F$, previously corresponding to the Hamiltonian change, see Eq.~(\ref{eq:Jarzynski}), is now itself a fluctuating function. Either it is $\Delta F = F\upo - F\upz$ or it is $\Delta F = F\upw - F\upz$ depending on the measurement outcome in each particular run. As a result the corresponding work fluctuation relation can be written as
\be \label{eq:feedbackJarzynski}
	\< e^{-\beta (W -  \Delta F)} \> = \gamma,
\ee 
where the average $\< \cdot \>$ includes an average over the two different protocols. 
Here $\gamma \ge 0 $ characterises the feedback efficacy, which is related to the reversibility of the non-equilibrium process. When $H(x, \lambda_1(\tau)) = H(x, \lambda_2(\tau))$, i.e. no feedback is actually implemented, then $\gamma$ becomes unity and the Jarzynski equality is recovered. The maximum value of $\gamma$ is determined by the number of different protocol options, e.g. for the two feedback choices discussed above one has $\gamma^{\max} = 2$. }

Apart from measuring $\< e^{-\beta (W -  \Delta F)} \>$ by performing the feedback protocol many times and recording the work done on the system, $\gamma$ can also be measured experimentally, see section \ref{sub:clexp}. To do so the system is initially prepared in a thermal state for one of the final Hamiltonians, say $H(y, \lambda_1(\tau))$, at inverse temperature $\beta$. The force parameter $\lambda_1(t)$ in the Hamiltonian is now changed \emph{backwards}, $\lambda_1(\tau) \to \lambda(0)$) without any feeback. The particle's final phase space point $y_{\tau}$ is recorded. The probability $P^{(1)}(y_{\tau})$ is established that the particle ended in the phase space volume that corresponds to the choice of $H(y, \lambda_1(\tau))$ in the forward protocol, in this example in the left box. The same is repeated for the other final Hamiltonians, i.e. $H(y, \lambda_2(\tau))$.  The efficacy is now the added probability for each of the protocols, $\gamma = \sum_k  P^{(k)} (y_{\tau})$. 

Applying Jensen's inequality to Eq.~(\ref{eq:feedbackJarzynski}) one now obtains $\< W \> \ge \<\Delta F\> - {1 \over \beta} \ln \gamma$, which when assuming that a cycle has been performed, $\<\Delta F\> = 0$, the maximal value of $\gamma$ for two feedback options has been reached, $\gamma=2$, and the inequality is saturated, becomes
\be
	\< W^{\rm demon}_{\rm ext} \> = - \< W \> =  k_B T \, \ln 2,
\ee
in agreement with Eq.~(\ref{eq:demon}). \ja{A recent review provides a detailed discussion of feedback in classical fluctuation theorems \cite{PHS}. The efficiency of feedback control in \emph{quantum} processes has also been analysed, see e.g. \cite{SU08, MT11, HJ14}.}

\subsection{Classical statistical physics experiments} \label{sub:clexp}

\begin{figure}[t]
 \includegraphics[width=0.37\textwidth]{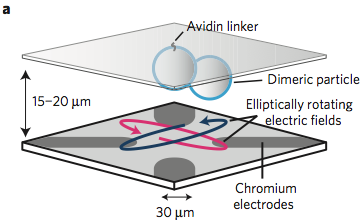}  
 \includegraphics[width=0.2\textwidth]{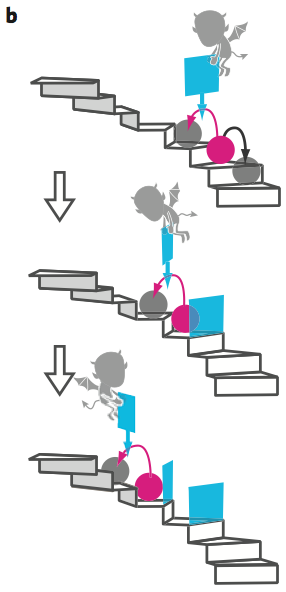}  \quad 
{\parbox[b]{0cm}{{\sf \footnotesize \bfseries c}\\\rule{0ex}{2.2in}}} \includegraphics[width=0.37\textwidth]{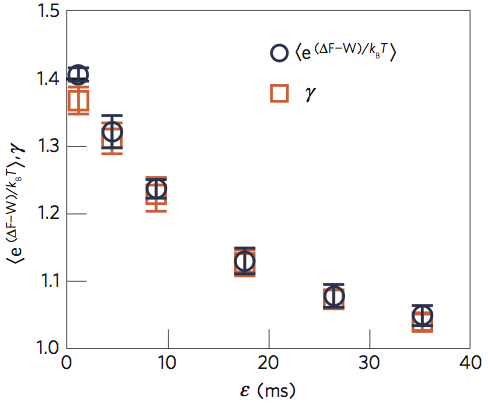} 
	\caption{\label{fig:Maxwell} {\sf  \bfseries  a}. Dimeric polystyrene bead and electrodes for creating the spiral staircase. {\sf  \bfseries  b}.  Schematic of a Brownian particle's dynamics climbing the stairs with the help of the demon. {\sf  \bfseries  c}. \ja{Experimental results of $\langle e^{-\beta (W - \Delta F)}\rangle$ for the feedback process, and of $\gamma$ for a time-reversed (non-feedback) process,} see Eq.~(\ref{eq:feedbackJarzynski}), over the demon's reaction time $\epsilon$.  \borrowedfig{(Figures taken from {\it Nature Physics} {\bf 6}, 988 (2010).)} }
\end{figure}

A large number of statistical physics experiments are  performed with colloidal particles, such as polystyrene beads, that are suspended in a viscous fluid. The link between information theory and thermodynamics has been confirmed by groundbreaking colloidal experiments only very recently. For example, the heat-to-work conversion achieved by a Maxwell demon, \ja{see subsections \ref{sub:demon} and \ref{sub:feedback},} intervening in the statistical dynamics of a system was investigated \cite{Toyabe10} with a dimeric polystyrene bead suspended in a fluid and undergoing rotational Brownian motion, see Fig.~\ref{fig:Maxwell}a. An external electrical potential was applied so that the bead was effectively moving on a spiral staircase, see Fig.~\ref{fig:Maxwell}b. The demon's action was realised by measuring if the particle had moved up and, depending on this, shifting the external ladder potential quickly, such that the particle's potential energy was ``saved'' and the particle would continue to climb up. The work done on the particle, $W$, was then measured for over 100000 trajectories to obtain an experimental value of $\< e^{-\beta (W - \Delta F)}\>$, required to be identical to 1 by the standard Jarzynski equality, Eq.~(\ref{eq:Jarzynski}). Due the feedback operated by the demon the value did turn out \emph{larger} than 1, see Fig.~\ref{fig:Maxwell}c, in agreement with Eq.~(\ref{eq:feedbackJarzynski}). A separate experiment implementing the time-reversed process without feedback was run to also determine the value of the efficacy parameter $\gamma$ predicted to be the same as $\langle e^{-\beta (W - \Delta F)}\rangle$. Fig.~\ref{fig:Maxwell}c shows very good agreement between the two experimental results for varying reaction time of the demon, reaching the highest feedback efficiency when the demon acted quickly on the knowledge of the particle's position. 
The theoretical predictions of fluctuation relations that include Maxwell's demon \cite{Sagawa-Ueda2010} have also been tested with a single electron box analogously to the original Szilard engine \cite{Koski15}.

\medskip

\begin{figure}[t]
\centering
	\includegraphics[width=0.42\textwidth]{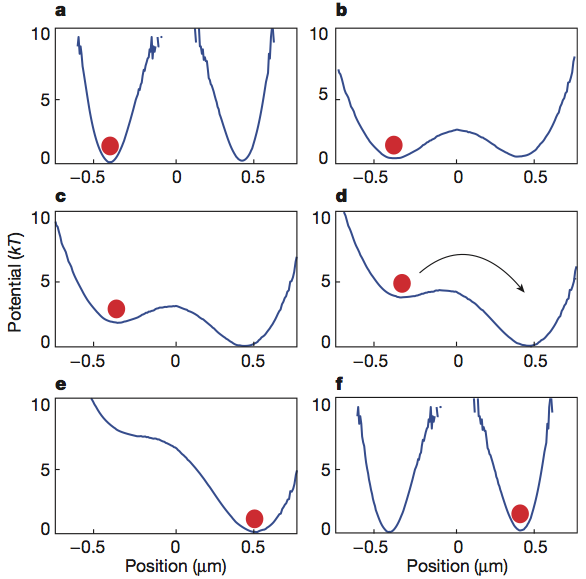} \quad \quad \quad
	\includegraphics[width=0.36\textwidth]{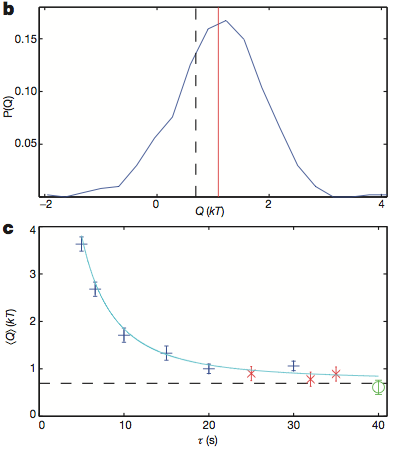} 
	\caption{\label{fig:Landauer}  {\sf  \bfseries Left a-f:} Erasure protocol for a particle trapped in a double well. Particles starting in either well will end up in the right hand side well at the end of the protocol. {\sf  \bfseries Right b:} Measured heat probability distribution $P(Q_{\rm dis})$ over fluctuating heat value $Q_{\rm dis}$ for a fixed protocol cycle time. {\sf  \bfseries Right c:} Average heat in units of $k_B T$, i.e. $\< Q_{\rm dis}\>/(k_B \, T)$, over varying cycle time $\tau$ in seconds. \borrowedfig{(Figures taken from {\it Nature} {\bf 483}, 187 (2012).)} 
	}
\end{figure}

Another recent experiment measured the heat dissipated in an erasure process, \ja{see subsection \ref{sub:erasure},} with a colloidal silica bead that was optically trapped with tweezers in a double well potential \cite{Berut12}. The protocol employed, see Fig.~\ref{fig:Landauer}, ensured that a bead starting in the left well would move to the right well, while a starting position in the right well remained unchanged. The lowering and raising of the energetic barrier between the wells was achieved by changing the trapping laser's intensity. The tilting of the potential, seen in Fig.~\ref{fig:Landauer}c-e, was neatly realised by letting the fluid flow, resulting in a force acting on the bead suspended in the fluid. The dissipated heat was measured by following the trajectories $x_t$ of the bead and integrating, $Q_{\rm dis} = - \int_0^{\tau} {\p {\cal U}(x_t, \lambda(t)) \over \p x_t} \, \dot{x}_t  \, \d t$, cf. Eq.~(\ref{eq:work+heat}).
\ja{Here} ${\cal U}(x_t, \lambda(t))$ is the explicitly time-varying potential that together with a constant kinetic term makes up a time-varying Hamiltonian $H (x_t, \lambda(t)) = T (x_t) + {\cal U}(x_t, \lambda(t))$. The measured heat distribution $P(Q_{\rm dis})$ and average heat $\< Q_{\rm dis}\>$, defined analogously to Eq.~(\ref{eq:trajW2}), are shown on the right in Fig.~\ref{fig:Landauer}. The time taken to implement the protocol is denoted by $\tau$. In the \ja{quasi-static limit,} i.e. the limit of long times in which the system equilibrates throughout its dynamics,  it was found that the Landauer limit, $k_B \, T \, \ln 2$, for the minimum dissipated heat is indeed approached. 

\medskip

\ja{Another beautiful high-precision experiment uses electrical feedback to effectively trap a colloidal particle and implement an erasure protocol \cite{Bech14}. This experiment provides a direct comparison between the measured dissipated heat for the erasure process as well as a non-erasure process showing that the heat dissipation is indeed a consequence of the erasure of information. Possible future implementations of erasure and work extraction processes using a \emph{quantum} optomechanical  system, consisting of a two-level system and a mechanical oscillator, have also been proposed \cite{ERA15}.}

\subsection{Definitions of quantum fluctuating work and heat} \label{sub:qfwh}

By its process character it is clear that work is not an observable \cite{Talkner2007}, i.e. there is no operator, $w$, such that $W= \tr[w \, \rho]$. To quantise the Jarzynski equality the crucial step taken is to define the fluctuating quantum work for \emph{closed} dynamics as a \emph{two-point correlation function}. I.e. a projective measurement of energy needs to be performed at the beginning and end of the process to gain knowledge of the system's energetic change. This enables the construction of a work-distribution function \ja{(known as the two-point measurement work distribution)} and allows the formulation of the Tasaki-Crooks relation and the quantum Jarzynski equality \cite{KURCHAN,TASAKI,Mukamel03,Talkner2007}. 

To introduce the quantum fluctuating work and heat, consider a quantum system with initial state $\rho\upz$ and initial Hamiltonian $H\upz  = \sum_n E\upz_n \, | e\upz_n \> \< e\upz_n |$ with eigenvalues $E\upz_n$ \ja{and energy eigenstates $| e\upz_n \>$.} A \emph{closed} system undergoes dynamics due to its time-varying Hamiltonian which generates a \ja{unitary transformation}\footnote{Here ${\cal T}$ stands for a time-ordered integral.}  $V\upa = {\cal T} e^{- i \int_0^{\tau} H\upt \, \d t / \hbar }$ and ends in a final state $\rho\upa = V\upa \, \rho\upz \, {V\upa}^{\dag}$ and a final Hamiltonian $H\upa = \sum_m E\upa_m \, | e\upa_m\> \< e\upa_m|$, with energies $E\upa_m$ \ja{and energy eigenstates $| e\upa_m \>$.} 
To obtain the fluctuating work the energy is measured at the beginning of the process, giving e.g. $E\upz_n$, and then again at the end of the process, giving e.g. in $E\upa_m$. The difference between the measured energies is now identified entirely with fluctuating work,
\be \label{eq:qW}
	W^{\close}_{m,n} := E\upa_m - E\upz_n,
\ee
as the system is closed, the evolution is unitary and no heat dissipation occurs, cf. the classical case (\ref{eq:zeroQ}). The work distribution is then peaked whenever the distribution variable $W$ coincides with $W^{\tau}_{m,n}$,
\be
	P(W) =  \sum_{n,m} p\upa_{n,m} \, \delta(W - (E\upa_m - E\upz_n)).
\ee
Here $p\upa_{n,m} = p\upz_n \, p\upa_{m|n}$ is the joint probability distribution of finding the initial energy level $E\upz_n$ and the final energy $E\upa_m$. This can be broken up into the probability of finding the initial energy $E\upz_n$, $p\upz_n =  \< e\upz_n | \rho\upz | e\upz_n \>$, and the conditional probability for transferring from $n$ at $t=0$ to $m$ at $t=\tau$,  $p\upa_{m|n}  = | \<e\upa_m | V\upa | e\upz_n \> |^2$.

The average work for the unitarily-driven non-equilibrium process can now be calculated as the average over the work probability distribution, i.e. using (\ref{eq:trajW2}),
\be
	\< W^{\close}_{\tau} \>
	&=& \int \sum_{n,m} p\upa_{n,m} \, \delta(W - (E\upa_m - E\upz_n)) \, W \, \d W \nonumber \\
	&=&  \sum_{n,m} p\upz_n \, p\upa_{m|n} \, (E\upa_m - E\upz_n) 
	=  \sum_{m} p\upa_m \, E\upa_m - \sum_{n} p\upz_n  \, E\upz_n,
\ee
where $p\upa_{m} :=  \sum_{n} p\upa_{n,m} = \< e\upa_m | \rho\upa | e\upa_m \>$ are the final state marginals of the joint probability distribution, i.e. just the probabilities for measuring the energies $E\upa_m$ in the final state $\rho\upa$. Thus the average work for the unitary process is just the internal energy difference, cf. (\ref{eq:closedW}),
\be	
	\< W^{\close}_{\tau} \> &=&  \tr[ H\upa  \, \rho\upa] - \tr[ H\upz \, \rho\upz] = \Delta U.
\ee

\subsection{Quantum fluctuation relations} \label{sub:QJarzynski}

With the above definitions, in particular that of the work distribution, the quantum Jarzynski equation can be readily formulated for a \emph{closed} quantum system undergoing externally driven (with unitary $V$) non-equilibrium dynamics \cite{KURCHAN,TASAKI,Mukamel03,Talkner2007,ESPOSITORMP}. The average exponentiated work done on a system starting in initial state $\rho\upz$ now becomes, cf.  (\ref{eq:Jarzynski}), 
\be	
	\< e^{- \beta W^{\close}_{\tau} }\>
	&=& \int P (W) \, e^{-\beta W} \, \d W 
	= \sum_{n,m} p\upa_{n,m} \, e^{-\beta (E\upa_m - E\upz_n)}.
\ee
When the initial state is a thermal state for the Hamiltonian $H^0$ at inverse temperature $\beta$, e.g. $\rho\upz = \sum_n p\upz_n  \, | e\upz_n \> \< e\upz_n |$, with thermal probabilities $p\upz_n = {e^{-\beta E\upz_n} \over Z\upz}$ and partition function $Z\upz = \sum_{n} \,  e^{-\beta E\upz_n}$, \ja{one obtains the quantum Jarzynski equality}
\be \label{eq:QJ}
	\< e^{- \beta W^{\close}_{\tau} }\>	
	&=& \ja{ \sum_{n,m} \, {e^{-\beta E\upz_n} \over Z\upz} \, p\upa_{m|n}  \, e^{-\beta (E\upa_m - E\upz_n)} }
	= {1 \over Z\upz} \, \sum_{m} \,  e^{-\beta E\upa_m} 
	= {Z\upa \over Z\upz} = e^{- \beta \Delta F}.
\ee
Here we have used that the conditional probabilities sum to unity, $\sum_n p\upa_{m|n} = \sum_n | \langle e\upa_m |  V\upa | e\upz_n \rangle|^2 = 1$ and $Z\upa =  \sum_{m} \,  e^{-\beta E\upa_m}$ is, like in the classical case, the partition function of the hypothetical thermal configuration for the final Hamiltonian, $H\upa$. The free energy difference is, like in the classical case, $\Delta F= - {1 \over \beta} \ln  {Z\upa \over Z\upz}$. 

\medskip

Similarly, the classical Crooks relation, cf. Eq.~(\ref{eq:Crooks}), can also be re-derived for the quantum regime and is known as the Tasaki-Crooks relation \cite{TASAKI},
\be \label{eq:TC}
  		{P^F(W) \over P^B(-W)} = e^{\beta(W-\Delta F)}.
\ee 

It is interesting to note that in the \ja{two-point measurement scheme} both, the quantum Jarzynski equality and the Tasaki-Crooks relation, show no difference to their classical counterparts, contrary to what one might expect. A debated issue is that the energy measurements remove coherences with respect to the energy basis  \cite{Talkner2007}, and these do not show up in the work distribution, $P(W)$. It has been suggested \cite{KA15} that the non-trivial work \ja{and heat that may be exchanged during the second measurement, see section \ref{sub:cohW}, implies that identifying the energy change entirely with fluctuating work (\ref{eq:qW}) is inconsistent. If the initial state is not thermal but has coherences, then also the first measurement will non-trivially affect these initial coherences and lead to work and heat contributions. Work probability distributions and generalised Jarzynski-type relations have been proposed to account for these coherences using path-integral and quantum jump approaches \cite{SG15,EAC15}.} 

\subsection{Quantum fluctuation experiments}

While a number of interesting avenues to test fluctuation theorems at the quantum scale have been proposed~\cite{Huber,Heyl,Pekola}, the experimental reconstruction of the work statistics for a quantum protocol has long remained elusive. A new measurement approach has recently been devised that is based on well-known interferometric schemes of the estimation of phases in quantum systems, which bypasses the necessity of two direct projective measurements of the system state \cite{Mazzola}. The first quantum fluctuation experiment that confirms the quantum Jarzynski relation, Eq.~(\ref{eq:QJ}), and the Tasaki-Crooks relation, Eq.~(\ref{eq:TC}), with impressive accuracy has recently been implemented in a Nuclear Magnetic Resonance (NMR) system using such an interferometric scheme \cite{batalhao}. Another, very recent experiment uses trapped ions to measure the quantum Jarzynski equality \cite{An15} with the two measurement method used to derive the theoretical result in section \ref{sub:QJarzynski}.

\begin{figure}[t]
\centering
	\includegraphics[width=0.38\textwidth]{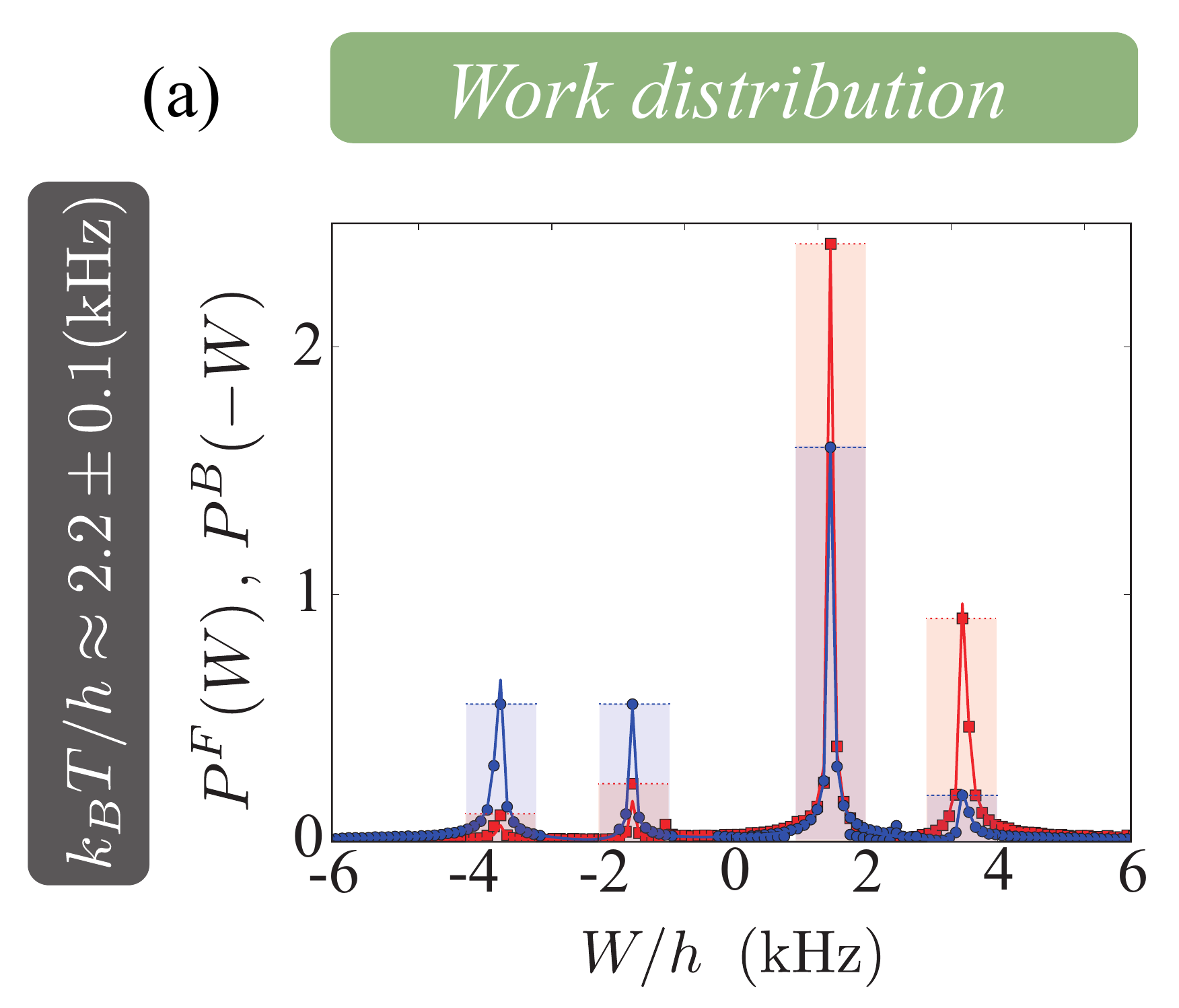} \quad 
	\includegraphics[width=0.58\textwidth]{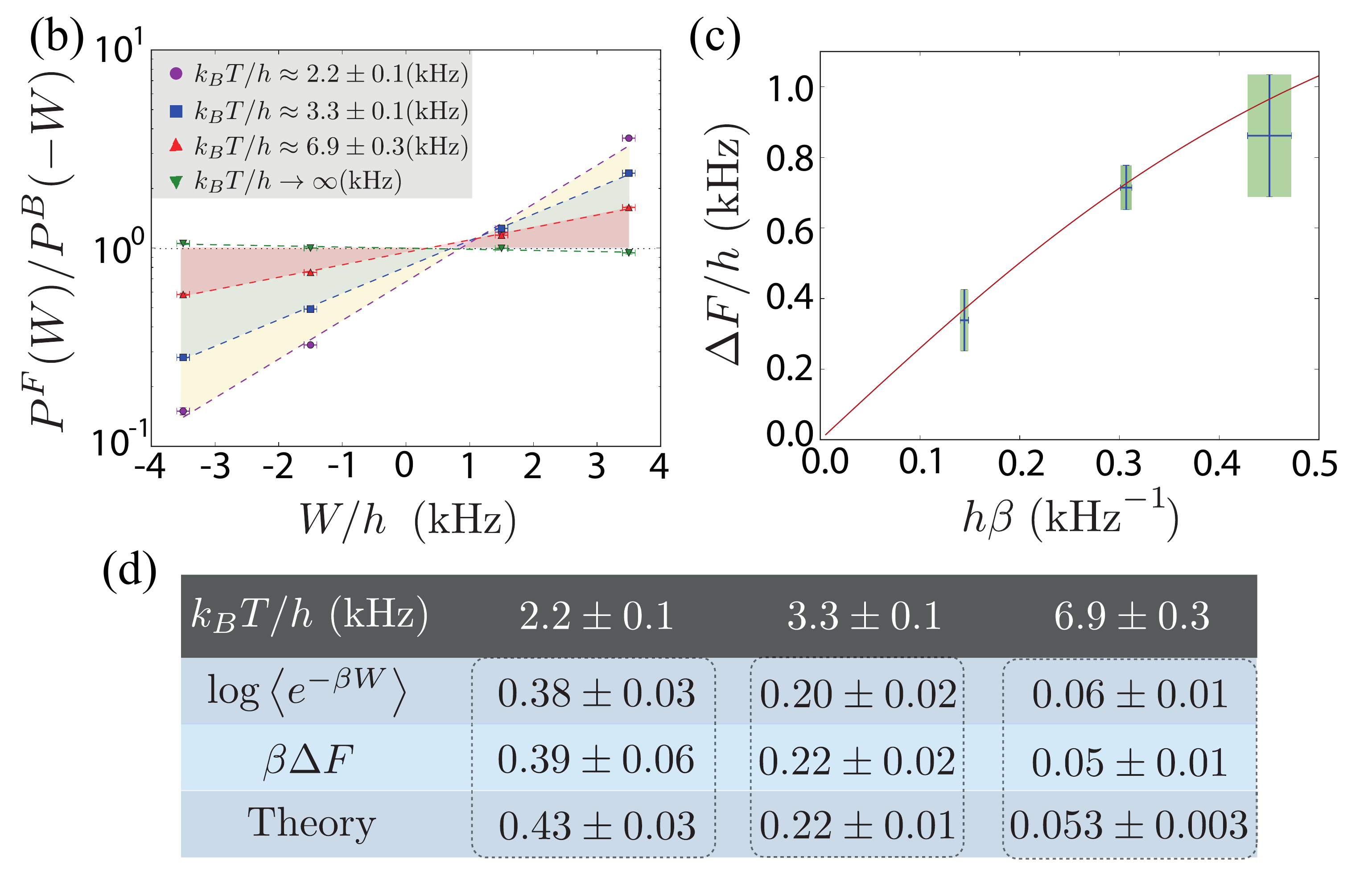}
	\caption{\label{fig:QJar}  {\sf \bfseries (a)}: Experimental work distributions corresponding to the forward (backward) protocol, $P^F(W)$ ($P^B(-W)$), are shown as red squares (blue circles). {\sf  \bfseries (b)} The Tasaki-Crooks ratio is plotted on a logarithmic scale for four values of the system's temperature. {\sf  \bfseries (c)} Crosses indicate mean values and uncertainties for $\Delta F$ and $\beta$ obtained from a linear fit to data compared with the theoretical prediction indicated by the red line. {\sf  \bfseries (d)} Experimental values of the left and right-hand side of the quantum Jarzynski relation measured for three temperatures together with their respective uncertainties, and theoretical predictions for $\ln {Z\upa \over Z\upz}$ showing good agreement. \borrowedfig{(Figures taken from {\it Phys. Rev. Lett.} {\bf 113}, 140601 (2014).)}
	}
\end{figure}

The NMR experiment was carried out using liquid-state NMR-spectroscopy of the $^{1}$H and $^{13}$C nuclear spins of a chloroform-molecule sample. This system can be regarded as a collection of identically prepared, non-interacting, spin-$1/2$ pairs~\cite{Ivan}. The $^{13}$C nuclear spin is the driven system, while the $^{1}$H nuclear spin embodies an \ja{auxiliary degree of freedom, referred to as ``ancilla''. Instead of performing two projective measurements on the system it is possible to reconstruct the system's work distribution with an interferometric scheme in which the ancilla is instrumental \cite{Mazzola}. The ancilla here interacts with the system at the beginning and the end of the process, and as a result its state acquires a phase. This phase corresponds to the energy difference experienced by the system and the ancilla is measured only once to obtain the energy difference of the system.} The $^{13}$C nuclear spin was prepared initially in a pseudo-equilibrium state $\rho\upz$ at inverse temperature $\beta$, for which four values were realised. The experiment implemented the time-varying spin Hamiltonian, $H\upt$, resulting in a unitary evolution, i.e. $V$ in section \ref{sub:QJarzynski}, by applying a time-modulated radio frequency field resonant with the $^{13}$C nuclear spin in a short time window $\tau$. The (forward) work distribution of the spin's evolution, $P^F(W)$ see Fig.~\ref{fig:QJar}(a), was then reconstructed through a series of one- and two-body operations on the system and the ancilla, and measurement of the transverse magnetization of the $^{1}$H nuclear spin. Similarly a backwards protocol was also implemented and the backwards work distribution, $P^B(W)$, measured, cf. sections \ref{sub:CFR} and \ref{sub:clexp}. 

The measured values of the Tasaki-Crooks ratio, $P^F(W)/P^B(-W)$, are shown over the work value $W$ in Fig.~\ref{fig:QJar}(b) for four values of temperature. The trend followed by the data associated with each temperature was in very good agreement with the expected linear relation confirming the predictions of the Tasaki-Crooks relation. The cutting point between the two work distributions was used to determine the value of $\Delta F$ experimentally which is shown in Fig.~\ref{fig:QJar}(c). Calculating the average exponentiated work with the measured work distribution, the data also showed good agreement with the quantum Jarzynski relation, Eq.~(\ref{eq:QJ}), see table in Fig.~\ref{fig:QJar}(d).

\section{Quantum Dynamics and Foundations of Thermodynamics}  \label{sec:qdft}
\add{The information theoretic approach to thermodynamics employs many standard tools of quantum information to study mesoscopic systems. For instance, an abstract view of dynamics, minimal in the details of Hamiltonians, is often employed in quantum information. Such a view of dynamics as a map between quantum states serves to produce rules common to generic dynamics and has served the study of computing and technologies well.} In this section, we describe some of these techniques and theorems and highlight how they are used in quantum thermodynamics.

\subsection{Completely positive maps}\label{sub:CPTP_sub}

We begin by reviewing the description of generic quantum maps \cite{kossakowski1972quantum,gorini1976completely,lindblad1976generators,gorini1978properties,breuer2002theory,nielsen2010quantum}. \add{The point of describing dynamics through a ``map" as opposed to a model of temporal dynamics (i.e., a Hamiltonian) is deliberate. Maps are not explicit functions of time (though they can be parametrized by time, as we will see later), but are two-point functions.}
\jan{They accept initial states of the dynamics they model and output final states. Specifically, a completely positive trace preserving (CPTP) map transforms input density matrices $\rho$ into \emph{physical} output states $\rho'$. The term ``physical'' here refers to the requirement that the output state is again a well-defined density matrix. 
A map $\Phi$ has to obey several rules to guarantee that its outputs are physical states.  These properties include:}
\begin{enumerate}
\item trace preservation: $\rho':=\Phi(\rho)$ has unit trace for all input states $\rho$ of dimensionality $d$. \add{If this is violated, the output states become unphysical in that Born's rule cannot be applied directly anymore.}
\item positivity: $\Phi(\rho)$ has non-negative eigenvalues, interpreted as probabilities.
\item complete positivity: \jan{$\forall k\in \{0,\ldots \infty\}: (\mathbb{I}^{(k)}\otimes\Phi ) \, (\sigma^{(k+d)})$ has non-negative eigenvalues for all states $\sigma^{(k+d)}$.} This subsumes positivity.
\end{enumerate}
The final property is the statement that if the CPTP map is acting on a subsystem of dimension $d$ which is part of a larger \sai{(perhaps entangled)} system \jan{of dimension $k+d$, whose state is $\sigma^{(k+d)}$}, then the resultant \jan{global state must also be a ``physical" state.}

\jan{For instance, let us consider the transpose map $\mathbb{T}$ which, when applied to a given density matrix, transposes all its matrix elements in a fixed basis. When applied to physical states of dimension $d$ this map always outputs physical states of dimension $d$. For example, for a general qubit state $\rho$ with Bloch-vector $\vec{r} = (x, y, z)$ the transposition map gives
\begin{align}
\rho = \frac{1}{2} \left(\begin{array}{cc} 
1+ z & x - \mathbbm{i} y \\
x + \mathbbm{i} y & 1 - z
\end{array} \right) 
\mapsto
\mathbbm{T}(\rho)= \frac{1}{2} \left(\begin{array}{cc} 
1+ z & x + \mathbbm{i} y \\
x - \mathbbm{i} y & 1 - z
\end{array} \right).
\end{align}
Note that the eigenvalues of the matrix are invariant under the transposition map $\mathbb{T}$.

We can now consider the action of the transpose map on a subsystem. For example, for the two-qubit Bell state $\vert\phi^{+}\rangle= (\vert00\rangle+\vert11\rangle)/\sqrt{2}$ one may apply transposition only on the second qubit. This map then gives
\begin{align}
\vert\phi^{+}\rangle \langle \phi^{+}\vert = 
\frac{1}{2} \left( \begin{array}{cccc}
1 & 0 & 0 & 1\\
0 & 0 & 0 & 0\\
0 & 0 & 0 & 0\\
1 & 0 & 0 & 1
\end{array} \right) 
\mapsto
\mathbb{I}\otimes\mathbb{T} \, \left(\vert\phi^{+}\rangle \langle \phi^{+}\vert \right)
= \frac{1}{2}\left( \begin{array}{cccc}
1 & 0 & 0 & 0\\
0 & 0 & 1 & 0\\
0 & 1 & 0 & 0\\
0 & 0 & 0 & 1
\end{array} \right),
\end{align}
where the resulting global state turns out to have one negative eigenvalue. Because of this property, transposition is not a completely positive map. This example demonstrates that positivity, e.g. maintaining positive eigenvalues of a single qubit state, does not imply complete positivity, e.g. maintaining positive eigenvalues of a two-qubit state when the map is only applied to one qubit. The fact the ``partial'' transposition is not positive when the initial state is entangled is used extensively as a criterion to detect entanglement \cite{nielsen2010quantum}.
}

CPTP maps are related to dynamical equations, which we discuss briefly. When the system is closed, and evolves under a unitary $V$, the evolution of the density matrix is given by $\rho'=V\rho V^{\dagger}$. A more general description of the transformations between states when the system is interacting with an external environment is given by Lindblad type master equations. Such dynamics preserves trace and positivity of the density matrix, while \sai{allowing the density matrix to vary otherwise}. Such equations have the general form
\begin{align}\label{lindblad_eqn}
\displaystyle\frac{\d \rho}{\d t}=-i[H,\rho]+\sum_{k}\left[ A_{k}\rho A_{k}^{\dagger}-\frac{1}{2}A_{k}^{\dagger}A_{k}\rho-\frac{1}{2}\rho A_{k}A_{k}^{\dagger}\right].
\end{align}
Here $A_k$ are Lindblad operators that describe the effect of the interaction between the system and the environment on the system's state. \add{Notice that if an initial state $\rho(t=0)$ is evolved to a final state $\rho(t=T)$, then the resulting transformation can be captured by a CPTP map $\Phi(\rho(t=0))=\rho(t=T)$.} Master equations are typically derived with many assumptions, including that the system is weakly coupled to the environment and that the environmental correlations decay sufficiently quickly such that the initial state of the system is uncorrelated to the environment. Lindblad type master equations have been extensively employed to the study of quantum engines, and this connection will be discussed in section \ref{study_engines}. Some master equations are not in the Lindblad form, see \cite{redfield1957theory} for example. We note that the structure of Eq.~(\ref{lindblad_eqn}) arises naturally if Markovianity, trace preservation of the density matrix and positivity are considered. To see that this should be the case, consider the general evolution step for the density matrix, which we write without loss of generality as $\d\rho=-(M\rho+\rho N)\d t$. From the Hermiticity of the density matrix, we have $\d\rho^{\dagger}=\d\rho$, implying $N=M^{\dagger}$. We can hence be tempted to write an evolution equation of the form $\d\rho=-(M\rho+\rho M^{\dagger})\d t$. If we write $M=iH+J$, this form does not preserve trace for the terms involving $J$. \sai{This problem is fixed by subtracting an appropriate term. If we write $J=F^{\dagger}F$, then we can write a trace preserving equation namely $\d\rho=-i[H,\rho]\d t+(2F\rho F^{\dagger}-F^{\dagger}F\rho-\rho F^{\dagger}F)\d t$, where $2F\rho F^{\dagger}$ ensures trace preservation. Writing the operator $F$ in an orthonormal basis and diagonalizing the resulting operator produces the general equation above. We note that care has to be taken that the dynamical equation also preserves positivity, see \cite{breuer2002theory} for a detailed derivation.}

\subsubsection*{\jan{Three theorems involving CPTP maps}} 
We now briefly mention three important theorems often used in the quantum information theoretic study of thermodynamics. \add{These theorems are the Stinespring dilation theorem, the channel-state duality theorem and the operator sum representation respectively. The importance of these theorems lies in the alternative perspectives they afford an abstract dynamical map. We mention the theorems briefly, alongside a brief remark about the light they shed on CPTP maps.}

\smallskip

The first of these, \jan{{\bf Stinespring dilation}} asserts that every CPTP map $\Phi$ can be built up from three fundamental operations, namely tensor product with an arbitrary environmental ancilla state, $\tau_E$, a joint unitary and a partial trace. This assertion is written as
\be \label{eq:stinespring}
	\rho'=\Phi(\rho)=\tr_E [V \, \rho\otimes\tau_E \, V^{\dagger}].
\ee
\jan{Stinespring dilation allows for any generic quantum dynamics of the system, $S$, to be thought of in terms of a global unitary $V$ acting on the system and an environment, $E$. If  the interactions between the system and the environment are known, then the time dependence of the map may be specified. For instance, if the Hamiltonian governing the joint dynamics is $H_{SE}$, then $V =V_{SE} (t):=\exp{(-iH_{SE}t)}$. However, Stinespring dilation allows the mathematical analysis of many properties of maps for general sets of unitaries. 
We note that many quantum thermodynamics papers assume $\tau_E$ to be the Gibbs state of the environment and use this to study, e.g., the influence of temperature on the dynamics of the quantum system which interacts with the environment.} For example, in \cite{hutter2012almost} the authors considered thermalisation of a quantum system\footnote{They prove a condition for the long time states of a system to be independent of the initial state of the system. This condition relies on smooth min- and max-entropies which are defined below.}, a topic we will discuss in section \ref{ETT}. 

\smallskip

Another theorem used in \cite{hutter2012almost} and other studies \cite{faist2015gibbs} of quantum thermodynamics is the \jan{{\bf channel-state duality}} or Choi-Jamio{\l}kowski isomorphism. It says that there is a state $\rho_\Phi$ isomorphic to every CPTP map $\Phi$, given by
\begin{align}
\rho_\Phi:=\mathbb{I}\otimes\Phi(\vert\varphi\rangle\langle\varphi\vert),
\end{align}
where \jan{$\vert\varphi\rangle= (1/ \sqrt{d}) \, \sum^d_{i}\vert i\rangle\otimes\vert i\rangle$} is a maximally entangled bipartite state. \add{The main purpose of this theorem is that it allows us to think of the CPTP map $\Phi$ as a state $\rho_{\Phi}$ which is often useful to make statements about the amount of correlations generated by a given CPTP map. For example, see \cite{vinjanampathy2014second} for a discussion on entropic inequalities that employ this formalism.}

\smallskip

The final theorem is the \jan{{\bf operator sum representation}}, which asserts that any map, which has the representation
\begin{align}
\displaystyle\rho'=\Phi(\rho):=\sum_{\mu}\mathbb{K}_{\mu} \, \rho \, \mathbb{K}^{\dagger}_{\mu},
\end{align}
with $\sum\mathbb{K}^{\dagger}_{\mu}\mathbb{K}_{\mu}=\mathbb{I}$ is completely positive (but not necessarily trace preserving). These operators, often known as Sudarshan-Kraus operators, are related to the ancilla state $\tau_E$ and $V$ in Eq.~(\ref{eq:stinespring}). We note that these theorems are often implicitly assumed while discussing quantum thermodynamics from the standpoint of CPTP maps, see \cite{PhysRevLett.110.230601,faist2015gibbs,binder2015quantum} for examples.

\subsubsection*{Entropy inequalities and majorisation}

We \jan{state key results from quantum information theory} relating to entropy often employed in studying thermodynamics. Specifically, we discuss convexity, subadditivity, contractivity and majorisation. We begin by defining quantum relative entropy as 
\begin{align}
S[\rho\Vert\sigma]:=\tr [\rho \ln \rho - \rho \ln \sigma ].
\end{align}
The negative of the first term is the von Neumann entropy of the state $\rho$, see section \ref{sub:1st+2ndlaw}. 

The relative entropy is jointly \jan{\emph{convex}} in both its inputs. This means that if $\rho=\sum_m p_m\rho_m$ and $\sigma=\sum_m p_m\sigma_m$, then 
\begin{align}
\displaystyle S\left[\rho\Vert\sigma\right]\leq\sum_kp_kS[\rho_k\Vert\sigma_k].
\end{align}
Furthermore, for any tripartite state $\rho_{ABC}$ defined over three physical systems labelled $A$, $B$ and $C$, the entropies of various marginals are bounded by the inequality
\begin{align}
S(\rho_{ABC})+S(\rho_{B})\leq S(\rho_{AB})+S(\rho_{BC}).
\end{align}
This inequality, known as the strong subadditivity inequality is equivalent to the joint convexity of quantum relative entropy and is used, for instance in \cite{wolf2008area}, to discuss area laws in quantum systems. 

The quantum relative entropy is monotonic, or \jan{\emph{contractive}}, under the application of CPTP maps. For any two density matrices $\rho$ and $\sigma$, this contractivity relation \cite{vedral2002role} is written as
\begin{align}
S[\rho\Vert\sigma]\geq S[\Phi(\rho)\Vert\Phi(\sigma)].
\end{align}
Contractivity is central to certain extensions of the second law to entropy production inequalities, see \cite{spohn1978entropy,sagawa2012second,vinjanampathy2014second,horowitz2014equivalent} and is also obeyed by trace distance. There has been recent interest in improving these inequality laws in relation to irreversibility of quantum dynamics \cite{petz1986sufficient,petz1988sufficiency,petz2003monotonicity,wilde2015recoverability}.

Finally, \jan{\emph{majorisation}} is a quasi-ordering relationship between two vectors. Consider two vectors $\vec{x}$ and $\vec{y}$, which are $n$-dimensional. Often in the context of quantum thermodynamics, these vectors will be the eigenvalues of density matrices. The vector $\vec{x}$ is said to majorise $\vec{y}$, written as $\vec{x}\succ \vec{y}$ if all partial sums of the ordered vector obey the inequality $\sum_{i}^{k\leq n}x^{\downarrow}_i\geq\sum_{i}^{k\leq n}y^{\downarrow}_i$. Here $\vec{x}^{\downarrow}$ is the vector $\vec{x}$, sorted in descending order. A related idea is the notion of  \jan{\emph{thermo-majorisation}} \cite{Horodecki2013}. Consider $p(E,g)$ to be the probability of a state $\rho$ to be in a state $g$ with energy $E$. At a temperature $\beta$, such a state is $\beta$-ordered if $e^{\beta E_k}p(E_k,g_k)\geq e^{\beta E_{k-1}}p(E_{k-1},g_{k-1})$, for $k=\{1,2,\ldots\}$. If two states $\rho$ and $\sigma$ are such that $\beta$-ordered $\rho$ majorises $\beta$-ordered $\sigma$, then the state $\rho$ can be transformed to $\sigma$ in the resource theoretic setting discussed in section \ref{sec:sst}. 

\subsubsection*{\jan{Smooth entropies}}

\jan{Smooth entropies are related to von Neumann entropy defined in (\ref{eq:vNE}) and the conditional entropy used in subsection \ref{sub:sideinfo}. Smooth entropies are central in the single shot and resource theoretic approaches to quantum thermodynamics, presented in section \ref{sec:sst}.} 

Let us consider a bipartite system, whose parties are labeled $A$ and $B$. The smooth entropies are defined in terms of min- and max-entropies conditioned on $B$, given by 
\be
	S_{\min} (A\vert B)_{\rho} 
	&:=&\sup_{\sigma_B} \, \sup \{ \lambda\in\mathbb{R}:2^{-\lambda}\mathbb{I}_{A}\otimes\sigma_B\geq\rho_{AB} \},\\
	S_{\max} (A\vert B)_{\rho}
	&:=& \sup_{\sigma_B} \, \ln\left[F(\rho_{AB},\mathbb{I}_A\otimes\sigma_B)\right]^2.
\ee
Here the supremum is over all states in the space of subsystem $B$, $\sigma_B$ and $F(x,y):=\Vert\sqrt{x}\sqrt{y}\Vert_1$, with $\Vert g\Vert_1:=\tr[\sqrt{g^{\dagger}g}]$ is the fidelity. \add{Note that we retain standard notation here, employed for instance in standard textbooks on quantum information theory \cite{wilde2013quantum}.} The smoothed versions of min-entropy, $S^{\varepsilon}_{\rm min}$ is defined as the maximum over the min-entropy $S_{\rm min}$ of all states $\sigma_{AB}$ which are in a radius (measured in terms of purified distance \cite{tomamichel2010duality}) of $\varepsilon$ from $\rho_{AB}$. Likewise, $S^{\varepsilon}_{\rm max}$ is defined as the minimum $S_{\rm max}$ over all states in the radius $\varepsilon$. This can be written as (see Chapter 4 in \cite{tomamichel2012framework})
\begin{align}
 S^{\epsilon}_\alpha(A)_{\rho}:= 
\begin{cases}
\displaystyle \min_{\tilde{\rho}}S_{\alpha}(A)_{\tilde{\rho}},& \text{if } \alpha < 1\\
\displaystyle \max_{\tilde{\rho}}S_{\alpha}(A)_{\tilde{\rho}},& \text{if } \alpha > 1
\end{cases}
,~0\leq\epsilon<1,
\end{align}
where $\tilde{\rho}\approx_{\varepsilon}\rho$ are an $\varepsilon$-ball of close states. We note that there is more than one smoothing procedure, and care needs to be taken to not confuse them \cite{hutter2012almost}.


\subsection{Role of fixed points in thermodynamics}
Fixed points of CPTP maps play an important role in quantum thermodynamics. This role is completely analogous to the role played by equilibrium states in equilibrium thermodynamics. In equilibrium thermodynamics, the equilibrium state is defined by the laws of thermodynamics. If two bodies are placed in contact with each other, and can exchange energy, they will eventually equilibrate to a unique state, see section \ref{ETT}. On the other hand, if the system of interest is not in thermal equilibrium, notions such as equilibrium, temperature and thermodynamic entropy are ill-defined. In the non-equilibrium setting, \cite{oono1998steady} proposed a thermodynamic framework where the equilibrium state was replaced by the fixed point of the map that generates the dynamics. If the steady state of a particular dynamics is not the thermal state, since there is entropy produced in the steady state, ``housekeeping heat" was introduced as a way of taking into account the deviation from equilibrium. This program, called ``steady-state thermodynamics" \cite{schlogl1971thermodynamics,oono1998steady,sasa2006steady}, for a classical system described by Langevin equation is described in \cite{sasa2006steady} and the connection to fluctuation theorems is elaborated.

For a quantum system driven by a Lindblad equation given in Eq.~(\ref{lindblad_eqn}), the fixed point is defined as the steady state of the evolution. Such a steady state is defined by the Hamiltonian and the Lindblad operators that describe the dynamics. If a system whose steady state is an equilibrium state is taken out of equilibrium, the system will relax back to equilibrium. This relaxation mechanism will be accompanied by a certain entropy production, which ceases once the system stops evolving (this can be an asymptotic process). In the non-equilibrium setting, let the given dynamics be described by a CPTP map $\Phi$. Such a map is a sufficient description of the dynamical processes of interest to us. Every such map $\Phi$ is guaranteed to have at least one fixed point \cite{granas2013fixed}, which we refer to as $\rho_{\star}$. If we compare the relative entropy $S[\rho\Vert\rho_{\star}]$ before and after the application of the map $\Phi$, then this quantity reduces monotonically with the application of the CPTP map $\Phi$. This is because monotonicity of relative entropy insists that $S[\rho_{1}\Vert\rho_{2}]\geq S[\Phi[\rho_{1}]\Vert \Phi[\rho_{2}]]$, $\forall\rho_1,\rho_2$. Since the entropy production ceases only when the state reaches the fixed point $\rho_{\star}$, we can compare an arbitrary initial state $\rho$ with the fixed point to study the deviation from steady state dynamics. Using this substitution, the monotonicity inequality can be rewritten as a difference of von Neumann entropies, namely
\begin{align}
S[\Phi[\rho]]-S[\rho]\geq-\tr\left[ \Phi[\rho] \, \ln \rho_\star - \rho \, \ln \rho_\star \right]
:=-\varsigma.
\end{align}
This result \cite{spohn1978entropy,yukawa2001second} states that the entropy production is bounded by the term on the right hand side, a factor which depends on the initial density matrix and the map. If $N$ maps $\Phi_i$ are applied in succession to an initial density matrix $\rho$, the total entropy change can be shown easily \cite{sagawa2012second} to be bounded by
 \begin{align}
\displaystyle S[\Phi[\rho]]-S[\rho]\geq -\sum_{k=1}^{N}\varsigma_k.
 \end{align}
Here $\Phi$ is the concatenation of the maps $\Phi:=\Phi_{N}\circ\Phi_{N-1}\circ\ldots\circ\Phi_{1}$. 
We refer the reader to \cite{zurek1994decoherence,SU08,sagawa2012second,esposito2010three,esposito2011second} for a discussion of the relationship between steady states and laws of thermodynamics and \cite{feldmann2004characteristics} for a discussion on limit cycles for quantum engines. In \cite{PhysRevE.92.032129}, the authors use the CP formalism described here to derive fluctuation relations discussed in Sec.~(\ref{sub:feedback}).


\subsection{Thermalisation of closed quantum systems}\label{ETT}

One of the fundamental problems in physics involves understanding the route from microscopic to macroscopic dynamics. Most microscopic descriptions of physical reality, for instance, are based on laws that are time-reversal invariant \cite{fano1996symmetries}. Such symmetries play a strong role in the construction of dynamical descriptions at the microscopic level but many problems persist when the derivation of thermodynamic laws is attempted from quantum mechanical laws. In this subsection, we highlight some examples of the problems considered and introduce the reader to some important ideas. We refer to three excellent reviews \cite{RevModPhys.80.885}, \cite{von2014some} and \cite{polkovnikov2011colloquium} for a detailed overview of the field. 

An information theoretic example of the foundational problems in deriving thermodynamics involves the logical paradox in reconciling the foundations of thermodynamics with those of quantum mechanics. Suppose we want to describe the universe, consisting of system $S$ and environment $E$. From the stand point of quantum mechanics, then we would assign this universe a pure state, commensurate with the (tautologically) isolated nature of the universe. This wave-function is constrained by the constant energy of the universe. On the other hand, if we thought of the universe as a closed system, from the standpoint of statistical mechanics, we assume the axiom of ``equal a priori probability" to hold. By this, we mean that each configuration commensurate with the energy constraint must be equally probable. Hence, we expect the universe to be in a maximally mixed state in the given energy shell. This can be thought of as a Bayesian \cite{jaynes2003probability} approach to states in a given energy shell. This is in contradiction with the assumption made about isolated quantum systems, namely that states of such isolated systems are pure states. In \cite{PSW06}, the authors resolved this by pointing out that entanglement is the key to understanding the resolution to this paradox. They replace the axiom of equal a priori probability with a provable statement known as the ``general canonical principle". This principle can be stated as follows: If the state of the universe is a pure state $\vert\phi\rangle$, then the reduced state of the system
\begin{align}
	\rho_S:=\tr_E [\vert\phi\rangle\langle\phi\vert]
\end{align}
is very close (in trace distance, see section \ref{sec:qdft}) to the state $\Omega_S$ for almost all choices of the pure state $\vert\phi\rangle$. The phrase ``almost all" can be quantified into the notion of typicality, see \cite{PSW06}. The state $\Omega_S$ is the canonical state, defined as $\Omega_S:=\tr_E(\mathcal{E})$, the partial reduction of the equiprobable state $\mathcal{E}$ corresponding to the maximally mixed state in a shell (or subspace) corresponding to a general restriction. \add{An example of such a restriction is the constant energy restriction, and hence the ``shell" corresponds to the subspace of states with the same energy.} Since this applies to any restriction, and not just the energy shell restriction we motivated our discussion with, the principle is known as the general canonical principle. 

The intuition here is that almost any pure state of the universe is consistent with the system of interest being close to the canonical state, and hence the axiom of equal a priori probability is modified to note that the system cannot tell the difference between the universe being pure or mixed for a sufficiently large universe. \add{We emphasize that though almost all pure states have a lot of entanglement across any cut (i.e., across any number of parties you choose in each subsystem and any number of subsystems you divide the universe into), not all states do. A simple example of this is product states of the form $\vert\psi_1\rangle\otimes\vert\psi_2\rangle\ldots\vert\psi_N\rangle$.} Entanglement between the subsystems is understood to play an important role in this kinematic (see below) description of thermalisation, leading to the effect that sufficiently small subsystems are unable to distinguish highly entangled states from maximally mixed states of the universe, see \cite{gemmer2003distribution}, \cite{goldstein2006canonical} and \cite{dubey2012approach} for related works. 

This issue of initial states is only one of the many topics of investigation in this modern study of equilibration and thermalisation \cite{linden2009quantum}. When thermalisation of a quantum system is considered, it is generally expected that the system will thermalise for any initial condition of the system (i.e., for ``almost all" initial states of the system, the system will end up in a thermal state) and that many (but perhaps not all) macroscopic observables will ``thermalise" (i.e., their expectation values will saturate to values predicted from thermal states, apart from infrequent recurrences and fluctuations). Let us consider this issue of the thermalisation of macroscopic observables in a closed quantum system. Let the initial state of the system be given by $\vert\phi\rangle=\sum_k c_k\vert \Psi_k\rangle$, where the $\vert \Psi_k\rangle$ are the energy eigenstates of the Hamiltonian, \jan{with energies assumed non-degenerate. The corresponding coefficients $c_k$ are non-zero only in a small band around a given energy $E_0$.} If we consider an observable $\mathrm{B}$, its expectation value in the time evolving state is given by 
\begin{align}
\langle\phi\vert e^{i H t} B e^{-i H t}\vert\phi\rangle:=\langle \mathrm{B}\rangle = \displaystyle\sum_k \vert c_k\vert^2 \mathrm{B}_{kk} +\sum_{km}c^{*}_mc_k  e^{i[E_m-E_k]t}\mathrm{B}_{km},
\end{align}
where $\mathrm{B}_{km}:=\langle \Psi_m\vert \mathrm{B}\vert \Psi_k\rangle$. The long time average of this expectation value is given by the first term  and is expected to thermalise if this represents a model of thermalisation. But from the axioms of the micro-canonical ensemble, the expectation value is expected to be proportional to $\sum_k \mathrm{B}_{kk}$, in keeping with equal ``a priori probability" about the mean energy $E_0$. This is a source of mystery, since the long time average $\sum_k \vert c_k\vert^2 \mathrm{B}_{kk}$ somehow needs to ``forget" about the initial state of the system $\vert\phi\rangle$, \sai{given by coefficients $c_k$}. 

This can happen in three ways.  The first mechanism is to demand that $\mathrm{B}_{kk}$ be non-zero in the narrow band discussed above and come outside the sum. This is called ``eigenstate thermalisation hypothesis" (ETH)\cite{srednicki1994chaos,rigol2008thermalisation}, \sv{see \cite{gogolin2015equilibration} for more details on ETH}. The second method to reconcile the long time average with the axiom of the micro-canonical ensemble is to demand that the coefficients $c_k$ can be constant and non-zero for a subset of indices $k$. The third mechanism for the independence of the long time average of expectation values of observables, and hence the observation of thermalisation in a closed quantum system is to demand that the coefficients $c_k$ be uncorrelated to $\mathrm{B}_{kk}$. This causes $c_k$ to uniformly sample $\mathrm{B}_{kk}$, the average given simply by the mean value, in agreement with the micro-canonical ensemble. ETH was numerically verified for hardcore bosons in \cite{rigol2008thermalisation} and we refer the interested reader to a review of the field in \cite{dunjko2012thermalisation}, \sv{see \cite{sivak2012thermodynamic,olshanii2015geometry} for a geometric approach to the discussion of thermodynamics and thermalisation.}

\sai{An integrable system is defined as a system where the number of conserved local quantities grows extensively. Here the notions of local and global are with respect to the constituent subsystems. We write local to differentiate quantities such as eigenstates of the Hamiltonian, which commute with the Hamiltonian but are usually global.} Consider the eigenvalues of such an integrable system, which are then perturbed with a non-integrable Hamiltonian perturbation. Since equilibration happens by a system exploring all possible transitions allowed between its various states, any restriction to the set of all allowed transitions will have an effect on equilibration.  Since there are a number of conserved \sv{local} quantities implied by the integrability of a quantum system, such a quantum system is not expected to thermalise. \sv{(We note that this is one of the many definitions of thermalisation and it has been criticised as being inadequate \cite{weigert1992problem}, see also section~9 of \cite{gogolin2015equilibration} for a detailed criticism of various definitions of thermalisation)}. On the other hand, the absence of integrability is no guarantee for thermalisation \cite{rigol2009breakdown}. Furthermore, if an integrable quantum system is perturbed so as to break integrability slightly, then the various invariants are assumed to approximately hold. Hence, the intuition is that there should be some gap (in the perturbation parameter) between the breaking of integrability and the onset of thermalisation. This phenomenon is called pre-thermalisation, and owing to the invariants, such systems which are perturbed out of integrability settle into a quasi-steady state \cite{barnett2011prethermalisation}. This is a result of the interplay between thermalisation and integrability \cite{kollar2011generalized} and was experimentally observed in \cite{gring2012relaxation}. 

\section{Single Shot Thermodynamics} \label{sec:sst}

\jan{The single shot regime refers to operating on a single quantum system, which can be a highly correlated system of many subsystems, rather than on an infinite ensemble of identical and independently distributed copies of a quantum system} often referred to as the i.i.d. regime \cite{nielsen2010quantum}. For long the issue of whether information theory and hence physics can be formulated reasonably in the single shot regime has been an important open question which has only been addressed in the last decade, see e.g. \cite{renner2008security}. 
An active programme of applying quantum information theory techniques, in particular the properties of CPTP maps discussed in section \ref{sec:qdft}, to thermodynamics is the extension of the laws and protocols applicable to thermodynamically large systems to ensembles of \emph{finite size} that in addition may host quantum properties. Finite size systems typically deviate from the assumptions made in the context of equilibrium thermodynamics and there has been a large interest in identifying limits of work extraction from such small quantum systems. 

\subsection{Work extraction in single shot thermodynamics} \label{sub:sst}

\begin{figure}[t]
\centering
	\includegraphics[width=0.38\textwidth]{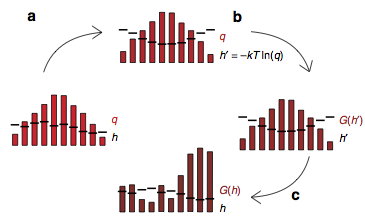} 
	\includegraphics[width=0.59\textwidth]{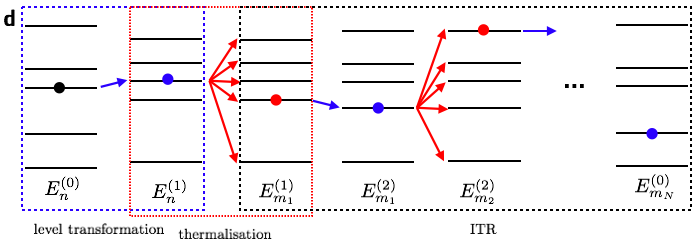} 
	\caption{\label{fig:Aberg}  
	\ja{Energy levels are indicated by the black lines, populations of levels are indicated as red columns.}
	{\sf \bfseries a:} \ja{Level transformations of the Hamiltonian changes only the energy levels while the populations of the levels remain the same. The energy levels can be chosen such that the initial probabilities correspond to thermal probabilities for a given temperature.}
	{\sf \bfseries b:}  ``Thermalisation'' of the population with respect to the energy levels resulting in the Gibbs state. 
	{\sf \bfseries c:} An isothermal reversible transformation (ITR) changing both the Hamiltonian's energy levels and the state population to remain thermal throughout the process. \borrowedfig{(Figures taken from {\it Nat. Commun.} {\bf 4,} 1925 (2013).)}   
	{\sf \bfseries d:} No level transfer is assumed during level transformations with the energy change  from level $E\upz_n$ corresponding to initial Hamiltonian $H=H\upz$ to $E\upo_n$ corresponding to changed Hamiltonian $H\upo$ associated with single shot work. Level transfers  during thermalisation are assumed to end with thermal probabilities in each of the energy eigenstates $E\upo_{m_1}$ of the same Hamiltonian $H\upo$, with the energy difference between $E\upo_n$ and $E\upo_{m_1}$ identified as single shot heat. An ITR is a sequence of infinitesimal small steps of LTs and thermalisations, here resulting in the system in an energy $E\upo_{m_N}$ after $N-1$ steps. }  
\end{figure}

One single shot thermodynamics approach to analyse how much work may be extracted from a system in a classical, diagonal distribution $\rho$ with Hamiltonian $H$ introduces \emph{level transformations} (LTs) of the Hamiltonian's energy eigenvalues while leaving the state unchanged and \emph{thermalisation} steps where the Hamiltonian is held fixed and the system equilibrates with a bath at temperature $T$ to a thermal state \cite{Aberg13}. \ja{Let the initial Hamiltonian be $H = \sum_{k, g_k} E\upz_k \, |E\upz_k, g_k\> \<E\upz_k, g_k|$ where $E\upz_k$ are the energy eigenvalues and $|E\upz_k, g_k\>$ the corresponding degenerate energy eigenstates, labelled with the degeneracy index $g_k$.
For the system starting with energy $E\upz_n$} the energy change during LTs, $E\upz_n \to E\upo_n$  see Fig.~\ref{fig:Aberg}d, is entirely associated with single shot work while the energy change, $E\upo_{m_1} \to E\upw_{m_2}$  see Fig.~\ref{fig:Aberg}d, for thermalisation steps is associated with single shot heat \cite{Aberg13}. See also the definition of average work and average heat in Eq.~(\ref{eq:Q}) associated with these processes for ensembles \cite{AG13}.
Through a sequence of LTs and thermalisations, see Fig.~\ref{fig:Aberg}a-c, it is found that the single shot random work yield of a system with diagonal distribution $\rho$ and Hamiltonian $H$ is \cite{Aberg13}
\be \label{eq:singleshotwork}
	W_{\rm yield} = k_B \, T \, \ln {r_n \over t_n}.
\ee 
\ja{Here  $r_n$ and $t_n$ are energy level populations of $\rho$ and the thermal state, $\tau := {e^{- \beta H} \over Z}$, respectively, which are both diagonal in the Hamiltonian's eigenbasis. The particular index $n$ appearing here refers to the initial energy state $|E\upz_n,g_n \>$ that the system happens to start with in a random single shot, see Fig.~\ref{fig:Aberg}d.}
Deterministic work extraction is rarely possible in the single shot setting and proofs typically allow a non-zero probability $\eps$ of failing to extract work in the construction of single shot work. The $\eps$-deterministic work content $A^{\eps} (\rho, H)$ is found to be \cite{Aberg13}  
\be  \label{eq:Aberg}
	A^{\eps} (\rho, H) \approx F^{\eps} (\rho, H) - F(\tau) := - k_B \, T \, \ln {Z_{\Lambda^* (\rho, H)} \over Z}, 
\ee
with the free energy difference defined through a ratio of \ja{a subspace partition function $Z_{\Lambda^*}$ over the full thermal partition function $Z$. Here $Z_{\Lambda^*} (\rho, H) := \sum_{(E,g) \in \Lambda^*} e^{- \beta E}$ is the partition function for the minimal subspace $\Lambda^* (\rho, H)  := \inf \{ \Lambda: \sum_{(E,g) \in \Lambda} \<E, g| \rho |E, g \> > 1 - \eps \}$ defined such that the state $\rho$'s probability in that subspace makes up the total required probability of $1-\eps$. 

As an example consider an initial state $\rho$ with probabilities $(5/15, 4/15, 1/15, 2/15, 3/15, 0)$ for non-degenerate energies $(0E, 1E, 2E, 3E, 4E, 5E)$ for $E>0$ and assume that the allowed error is $\epsilon = 1/10$. The thermal partition function is $Z = e^{- \beta 0} + e^{- \beta 1E} + e^{- \beta 2E} + e^{- \beta 3E} + e^{- \beta 4E} + e^{- \beta 5E}$. The subspace partition function is then $Z_{\Lambda^*} = e^{- \beta 0} + e^{- \beta 1E} + e^{- \beta 3E} + e^{- \beta 4E}$ including the thermal terms for the most populated energies until the corresponding state probability is at least $1 - 1/10$, i.e. $p_{0E, 1E, 3E, 4E} = 14/15 \ge 1 - 1/10$. 

The proof of Eq.~(\ref{eq:Aberg}) employs a variation of Crook's relation, see section \ref{sub:CFR}, and relies on the notion of smooth entropies, see section \ref{sub:CPTP_sub}. Contrasting with the ensemble situation that allows probabilistic fluctuations $P(W)$ around the average work, see e.g. Eq.~(\ref{eq:trajW2}), the single shot work $A^{\eps} (\rho, H)$ discussed here is interpreted as the amount of \emph{ordered energy}, i.e. energy with no fluctuations, that can be extracted from a distribution in a single-shot setting. }


\subsection{Work extraction and work of formation in thermodynamic resource theory} \label{sub:ho}

An alternative single shot approach is the \emph{resource theory} approach to quantum thermodynamics \cite{RM76}, \ja{where recent contributions include references }\cite{Janzing,Brandao08,resource,Horodecki2013,Aberg13,Brandao13b,SSP14,Lostaglio15}. Resource theories in quantum information identify a set of restrictive operations that can act on ``valuable resource states''. For a given initial state these restrictive operations then define a set of states that are reachable. For example, applying stochastic local operations assisted by classical communication (SLOCC) on an initial product state of two parties will produce a restricted, separable set of two-party states. The same SLOCC operations applied to a two party Bell state will result in the entire state space for two parties, thus the Bell state is a ``valuable resource''. In the  thermodynamic resource theory the valuable resource states are non-equilibrium states while the restrictive operations include thermal states of auxiliary systems.

\begin{figure}[t]
\centering
	\includegraphics[width=0.45\textwidth]{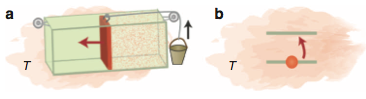}
	\quad \quad \quad
	\includegraphics[width=0.35\textwidth]{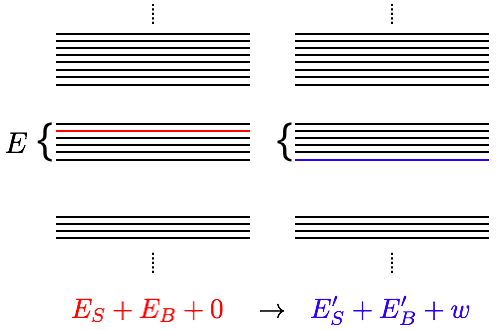} 
	\caption{\label{fig:workbit} 
	{\sf \bfseries Left a:} Work in classical thermodynamics is identified with lifting a bucket against gravity thus raising its potential energy.  {\sf \bfseries Left b:} In the quantum thermodynamic resource theory a work bit, \ja{i.e. a quantum version of a bucket,} is introduced where the jump from one energetic level to another, defines the extracted work during the process. \ja{In the text the lower level is denoted as $|0\>_W$, i.e. the ground state with energy $0$ and the excited state is $|w\>_W$ with energy $w$.}  \borrowedfig{(Figure taken from {\it Nature Comm.} {\bf 4} 2059 (2013).)} 
	{\sf \bfseries Right:} Global energy levels of combined system, bath and work storage system have  a high degeneracy indicated by the sets of levels all in the same energy shell, $E$. The resource theory approach to work extraction \cite{Horodecki2013} allows a global system starting in a particular energy state made up of energies $E =E_S+E_B+0$, indicated in red, to be moved to a final state made up of energies $E =E'_S+E'_B+w$, indicated in blue, in the same energy shell, $E$. 	Work extraction now becomes an optimisation problem to squeeze the population of the initial state into the final state and find the maximum possible $w$ \cite{GA15}.
	}
\end{figure}

The thermodynamic resource theory setting involves three components: the system of interest, $S$, a bath $B$, and a weight (or work storage system or battery) $W$. Additional auxiliary systems may also be included \cite{Brandao13b}  to explicitly model the energy exchange that causes the time-dependence of the system Hamiltonian, cf. section \ref{sec:cqnesp}, and to model catalytic participants in thermal operations, see section \ref{sub:2ndlaws}. In the simplest case, the Hamiltonian at the start and the end of the process is assumed to be the same and the sum of the three local terms, $H= H_S + H_B + H_W$. 
\emph{Thermal operations} are those transformations of the system that can be generated by a global unitary, $V$, that acts on system, bath and work storage system initially in a product state $\rho_S \otimes \tau_B \otimes |E_W^\ini \>_W\< E_W^\ini|$, where \ja{the work storage system starts in one of its energy eigenstates, $|E_W^\ini \>$, and the bath in a thermal state, $\tau_B =  \frac{e^{-\beta H_B}}{Z_B}$, which is considered a ``free resource''. The non-trivial resource for the process is the non-equilibrium state of the system, $\rho_S$.} Perfect energy conservation is imposed by requiring that the unitary may only induce transitions within energy shells of total energy $E =E_S+E_B+E_W$, see Fig.~\ref{fig:workbit}b. This is equivalent to requiring that $V$ must commute with the sum of the three local Hamiltonians. 

A key result in this setting is the identification of a maximal work \cite{Horodecki2013} that can be extracted from a single system starting in a diagonal non-equilibrium state $\rho_S$ to the work storage system under thermal operations. \ja{The desired thermal transformation, $\rho_{S} \otimes \tau_B \otimes |0\>_W\<0| \stackrel{V}{\longrightarrow} \sigma_{SB} \otimes |w\>_W\<w|$, enables the lifting of the work storage system by an energy $w>0$ from the ground state $|E_W^\ini \> = |0 \>_W$ to another single energy level $|w\>_W$, see Fig.~\ref{fig:workbit}a. 
Here the system and bath start in a product state and may end in a correlated state $\sigma_{SB}$.} The perfect transformation can be relaxed by requesting the transformation to be successful with probability $1 - \eps$, see section \ref{sub:sst}. The maximum extractable work in this setting is then \cite{Horodecki2013, GA15} 
\be \label{eq:ho}
	w^{\max}_\eps 
	= -\frac{1}{\beta} \,  \ln \sum_{E_S, g_S} \, t_{(E_S,g_S)} \, h_{\rho_S}( E_S, g_S, \eps)
	= : F^{\min}_\eps (\rho_S) - F(\tau_S),
\ee
where $\tau_S$ is the thermal state of the system at the bath's inverse temperature $\beta$ with eigenvalues $ t_{(E_S,g_S)} = \frac{e^{- \beta E_S}}{Z_S}$ and partition function $Z_S = \sum_{E_S, g_S} e^{- \beta E_S}$. Here $E_S$ and $g_S$ refer to the energy and degeneracy index of the eigenstates $|E_S, g_S \>$ of the Hamiltonian $H_S$. $h_{\rho_S}( E_S, g_S, \eps)$ is a binary function that leads to either including or excluding a particular $t_{(E_S,g_S)}$ in the summation, see \cite{Horodecki2013, GA15} for details.
\ja{Comparing with the result in Eq.~(\ref{eq:Aberg}) derived with the single shot approach discussed above one notices that the ratio ${Z_{\Lambda^*} / Z}$ is just the sum given in (\ref{eq:ho}), i.e. ${Z_{\Lambda^*} / Z} = \sum_{E_S, g_S} \, t_{(E_S,g_S)} \, h_{\rho_S}( E_S, g_S, \eps)$ and the expressions for the extractable work coincide. }

\ja{For non-zero $\eps$ the derivation can no longer rely on the picture of lifting the work storage system from its ground state to a pure excited state. A more general mixed work storage system state is used in \cite{SW15} and a general link between the uncertainty arising from the $\eps$-probabilities and the fluctuations in work as discussed in subsection \ref{sub:qfwh} is derived. This approach recovers the results Eq.~(\ref{eq:Aberg}) and (\ref{eq:ho}) as the lower bound on the extractable work.}
When extending the single level transitions of the work storage system to multiple levels \cite{GA15} it becomes apparent that this resource theory work cannot directly be compared to lifting a weight to a certain height \emph{or above}. Allowing the weight to rise to an energy level $|w\>_W$ or a range of higher energy levels turns the problem into a new optimisation and that would have a different associated ``work value'', $w'$, contrary to physical intuition. Due to its restriction to a particular energetic transition the above resource theory work may very well be a useful concept for \emph{resonance processes} where such a restriction is crucial \cite{GA15}. 

Similar to the asymmetry between distillable entanglement and entanglement of formation it is also possible to derive a \emph{work of formation} to create a diagonal non-equilibrium state $\rho_S$ which is in general smaller than the maximum extractable work derived above \cite{Horodecki2013},
\be
	w^{\min \, \rm (formation)}_\eps = {1 \over \beta}  \inf_{\rho_S^{\eps}} \, \left( \ln \, \min \{ \lambda: \rho_S^{\eps} \le \lambda \tau_S \} \right),
\ee
where $\rho_S^{\eps}$ are states close to $\rho_S$, $||\rho_S^{\eps} - \rho_S|| \le \eps$ with $|| \cdot ||$ the trace norm.

\subsection{Single shot second laws} \label{sub:2ndlaws}

Beyond optimising work extraction, a recent paper identifies the set of all states that can be reached in a single shot by a broader class of thermal operations in the resource theory setting \cite{Brandao13b}. This broader class of operations, ${\cal E}_C$, is called \emph{catalytic thermal operations} and involves, in addition to a system $S$ (which here may include the work storage system) and a heat bath, $B$, a catalyst $C$ \cite{JP99}. The role of the catalyst is to participate in the operation while starting and being returned uncorrelated to system and bath, and in the same state $\sigma_C$,
\be \label{eq:catalysis}
	{\cal E}_C (\rho_S) \otimes \sigma_C := \tr_{B}[V \, (\rho_S \otimes \tau_B \otimes \sigma_C) \, V^{\dag}].
\ee
The unitary now must commute \ja{with the sum} of the three Hamiltonians that the system, bath and catalyst start and end with, $[V, H_S+H_B+H_C]=0$. \ja{Note, that the map ${\cal E}_C$ on the system state depends upon the catalyst state $\sigma_C$.} For system states $\rho_S$ diagonal in the energy-basis of $H_S$ it is shown \cite{Brandao13b} that a state transfer under catalytic thermal operations is possible only under a \emph{family of necessary and sufficient conditions},
\be	\label{eq:2ndlaws}
	\rho_S \to \rho'_S = {\cal E}_C (\rho_S) \quad 
	\Leftrightarrow \quad
	\forall \alpha \ge 0:	\Fa(\rho_S, \tau_S) \ge \Fa(\rho'_S, \tau_S),
\ee
where $\tau_S$ is the thermal state of the system. $\tau_S$ is the fixed point of the map ${\cal E}_C$ and the proof makes use of the contractivity of map, discussed in section \ref{sub:CPTP_sub}. 
Here $\Fa(\rho_S,\tau_S) := k_B \, T \, \left( \Ren(\rho_S||\tau_S) - \ln Z_S \right)$ is a family of free energies defined through the $\alpha$-relative Renyi entropies, $\Ren(\rho_S||\tau_S): = {\mbox{sgn}(\alpha) \over  \alpha -1} \ln \sum_n r_n^{\alpha} \, t_n^{1-\alpha}$, applicable for states diagonal in the same basis. The $r_n$ and $t_n$ are the eigenvalues of $\rho_S$ and $\tau_S$ corresponding to energy eigenstates $|E_n, g_n \>$ of the system, cf. Eq.~(\ref{eq:singleshotwork}). An extension of these laws to non-diagonal system states and including changes of the Hamiltonian is also discussed \cite{Brandao13b}.
 
The resulting continuous family of second laws can be understood as limiting state transfer in the single shot regime in a \emph{more restrictive} way than the limitation enforced by the standard second law of thermodynamics, Eq.~(\ref{eq:Clausius}), valid for macroscopic ensembles \cite{Brandao13b}, see section \ref{sec:qt}. Indeed, the latter is included in the family of second laws: for $\lim_{\alpha \to 1}$ where $\Ren(\rho_S||\tau_S) \to \Reno(\rho_S||\tau_S) = \sum_n r_n \ln {r_n \over t_n}$ \cite{tomamichel2012framework} one recovers from Eq.~(\ref{eq:2ndlaws}) the standard second law, 
\be  
	k_B \, T \, \left(\sum_n r_n \ln {r_n \over t_n} - \ln Z_S \right) 
	&\ge& k_B \, T \, \left(\sum_n r'_n \ln {r'_n \over t_n} - \ln Z_S \right), \\
	\sum_n r_n \ln r_n - \sum_n r'_n  \ln r'_n 
	&\ge& - \sum_n (r'_n - r_n) \ln t_n, \\
	\Delta S 
	&\ge& \beta  \Delta U = \beta  \< Q \>,
\ee
with no work contribution, $\<W \> =0$, as no explicit source of work was separated here (although this can be done, see section \ref{sub:ho}). \ja{Moreover, in the limit of large ensembles and for systems that are not highly correlated the Renyi free energies reduce to the standard Helmholtz free energy, $\Fa(\rho_S, \tau_S) \approx F_1(\rho_S, \tau_S)$ for all $\alpha$, providing a single-shot explanation of why Eq.~(\ref{eq:Clausius}) is the only second law relation for macroscopic ensembles.} 



\ja{Allowing the catalyst in Eq.~(\ref{eq:catalysis}) to be returned after the operation not in the identical state, but a state close in trace distance, leads to the rather unphysical puzzle of thermal embezzling: a large set of transformations are allowed without the usual second law restrictions. This puzzle has been tamed by recent results showing that when physical constraints, such as fixing the dimension and the energy level structure of the catalyst, are incorporated the allowed state transformations are significantly restricted \cite{Ng15}.}

\subsection{Single shot second laws with coherence} \label{sub:coh2ndlaws}

\begin{figure}[t]
\centering
	\includegraphics[width=0.42\textwidth]{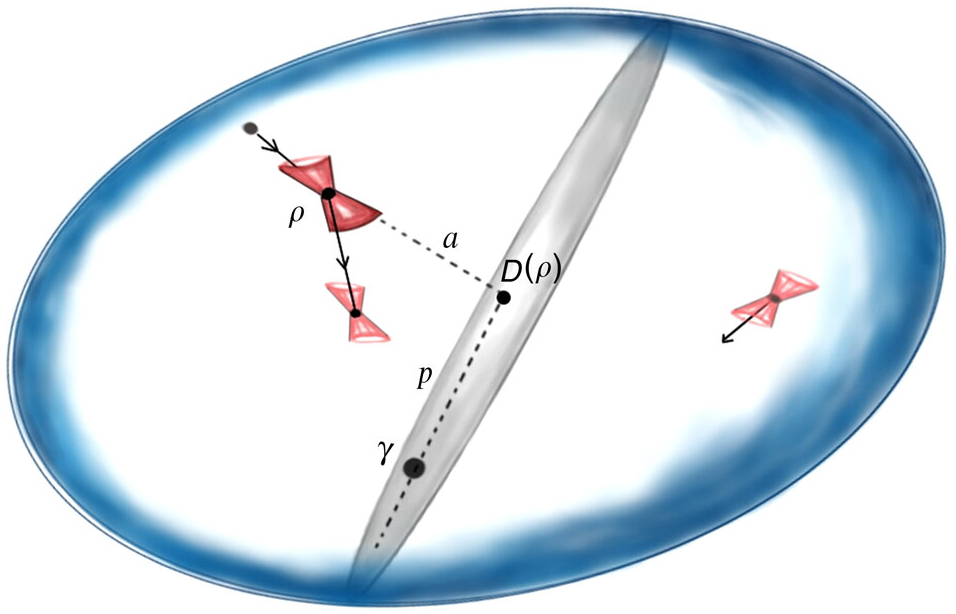} 
	\caption{\label{fig:2ndlaws}  
	For a fixed Hamiltonian $H_S$ and temperature $T$,  the manifold of time-symmetric states, indicated in grey, including the thermal state, $\gamma \equiv \tau_S$, as well as any diagonal state, $D (\rho) \equiv {\cal D}_{H_S} (\rho_S)$, is a submanifold of the space of all system states (blue oval). Under thermal operations an initial state $\rho \equiv \rho_S$ moves towards the thermal state  in two independent ``directions''. The thermodynamic purity $p$ measured by $\Fa$, and the asymmetry $a$ measured by $A_{\alpha}$, must both decrease. 
		\borrowedfig{(Figure taken from {\it Nat. Commun.} {\bf 6,} 6383 (2015).)} 
		}
\end{figure}

Recently it was discovered that thermal operations, ${\cal E} (\rho_S) := \tr_{B}[V \, (\rho_S \otimes \tau_B) \, V^{\dag}]$, on an initial state $\rho_S$ that has \emph{coherences} \ja{(i.e. non-zero off-diagonals) with respect to the energy eigenbasis of the Hamiltonian $H_S$}, require a second family of inequalities to be satisfied \cite{Lostaglio15}, in addition to the free energy relations for diagonal states (\ref{eq:2ndlaws}). The derived \ja{ \emph{necessary}} conditions are
\be \label{eq:coh2ndlaws}
	\rho_S \to \rho'_S = {\cal E} (\rho_S) \quad 
	\Rightarrow  \quad
	\forall \alpha \ge 0:	A_{\alpha}(\rho_S) \ge A_{\alpha}(\rho'_S),
\ee
where $A_{\alpha}(\rho_S) = \Ren(\rho_S || {\cal D}_{H_S} (\rho_S))$ and ${\cal D}_{H_S} (\rho_S)$ is the operation that removes all coherences between energy eigenspaces, i.e. ${\cal D}_{H} (\rho) \equiv \eta$ in section \ref{sub:cohW}. The $A_{\alpha}$ are coherence measures, i.e. they are coherence monotones for thermal operations. The proof of (\ref{eq:coh2ndlaws}) relies again on the contractivity \ja{of the relative entropy with respect to} the map ${\cal E}$, see section \ref{sub:CPTP_sub}, and its commuting with ${\cal D}$. \ja{It is noted that (\ref{eq:coh2ndlaws}) provides necessary conditions only - it is not known if the conditions on the $A_{\alpha} (\rho'_S)$ are sufficient for a thermal operation to exist that turns $\rho_S$ into $\rho'_S$.} 

The conditions (\ref{eq:coh2ndlaws}) mean that thermal operations on a single copy of a \emph{quantum} state with coherence must decrease its coherence, just like free energy must decrease for diagonal non-equilibrium states. The two families of second laws (\ref{eq:2ndlaws}) and (\ref{eq:coh2ndlaws}) can be interpreted geometrically as moving states in state space under thermal operations ``closer''\footnote{The Renyi divergences $\Ren$ are not proper distance measures \cite{tomamichel2012framework}.} to the thermal state in two independent ``directions'', in thermodynamic purity and time-asymmetry \cite{Lostaglio15}, see Fig.~\ref{fig:2ndlaws}. The decrease of coherence implies a quantum aspect to irreversibility in the resource theory setting \cite{LKJR15}. 
\ja{Again, when taking the single shot results to the macroscopic limit, $\rho \to \rho^{\otimes N}$, conditions (\ref{eq:coh2ndlaws}) become trivial and the standard second law (\ref{eq:Clausius}) is recovered as the only constraint. }

\section{Quantum Thermal Machines } \label{sec:Tmachines}
\add{Until now, we have discussed fundamental issues in quantum thermodynamics. Thermodynamics is a field whose focus from the very beginning has been both fundamental as well as applied. Applications of thermodynamics are mostly in designing thermal machines, which extract useful work from thermal baths. Examples of this are engines and refrigerators. \sai{In Sec~2, we discussed work that can be extracted from correlations and from coherences in the energy eigenbasis.} These coherences and correlations hence affect cyclical and non-cyclical machine operations and contribute towards the quantum features of thermal machine design. Here, we will present some of the recent progress in the design of quantum thermal machines that apply these aforementioned principles, which generalise classical machines. We will focus on two main types of quantum thermal machines. The first are cyclical machines which are quantum generalisations of engines and refrigerators. The second type are non-cyclical machines, which highlight important aspects of quantum thermal machines such as work extraction, power generation and correlations between subsystems.}

\subsection{Quantum Thermal Machines}
The study of the efficient conversion of various forms of energy to mechanical energy has been a topic of interest for more than a century. Quantum thermal machines (QTMs), defined as quantum machines \sai{that convert heat to useful forms of work} have been a topic of intense study. Engines, heat driven refrigerators, power driven refrigerators and several other kinds of thermal machines have been studied in the quantum regime, initially motivated from areas such as quantum optics \cite{scovil1959three,geusic1967quantum}. QTMs can be classified in various ways. Dynamically, QTMs can be classified as those that operate in discrete strokes, and those that operate as continuous devices, where the steady state of the continuous time device operates as a QTM. Furthermore, such continuous engines can be further classified as those that employ linear response techniques \cite{mazza2014thermoelectric} and those that employ dynamical equation techniques \cite{kosloff2014quantum}. These models of quantum engines are in contrast with biological motors, that extract work from fluctuations and have been studied extensively \cite{hanggi2009artificial}.

Engines use a working fluid and two (or more) reservoirs to transform heat to work. These reservoirs model the bath that inputs energy into the working fluid (hot bath) and another that accepts energy from the working fluid (cold bath). Two main classes of quantum systems have been studied as working fluids, namely discrete quantum systems and continuous variable quantum systems \sai{\cite{kosloff1984quantum,lin2003performance,quan2007quantum,zheng2014work}. This study of continuous variable systems} complements various models of finite level systems \cite{geva1992quantum,zhang2007four} and hybrid models \cite{youssef2009quantum,youssef2010quantum} studied as quantum engines. Both continuous variable models of engines and finite dimensional heat engines have been theoretically shown to be operable at theoretical maximal efficiencies. Such efficiencies are only well defined once the engine operation is specified. 

The efficiency of any engine is given by the ratio of the net work done by of the system to the heat that flows into the system, namely
\begin{align}
	\eta=\frac{\<W_{\rm net}\>}{\< Q_{\rm in}\>}.
\end{align}
From standard thermodynamics, it is well known that the most efficient engines also output zero power \sai{since they are quasistatic}. Hence, for the operation of a real engine, other objective functions such as power have to be optimised. The seminal paper by Curzon and Ahlborn \cite{curzon1975efficiency} considered the issue of the efficiency of a classical Carnot engine operating between two temperatures, cold $T_C$ and hot $T_H$, at maximum power, which they found to be
\be \label{eq:CA}
	\eta_{\rm CA} = 1 - \sqrt{\frac{T_{\rm C}}{T_{\rm H}}}.
\ee
This limit is also reproduced with quantum working fluids, as discussed in section \ref{sub:CWE}. We note that the aforementioned formula for the efficiency at maximum power is sensitive to the constraints on the system, and changing the constraints changes this formula. For instance, see \cite{uzdin2014universal,schmiedl2008efficiency} for results generalising the Curzon-Ahlborn efficiency to stochastic thermodynamics and to quantum systems with other constraints.

The rest of this section is a brief description of how engines are designed in the quantum regime. We consider three examples, namely Carnot engines with spins as the working fluid, harmonic oscillator Otto engines and Diesel engines operating with a particle in a box as the working fluid. \sai{These examples were chosen to illustrate the design of a variety of engines operating with a variety of quantum working fluids.}
\subsection{Carnot Engine}\label{study_engines}

Classically, the Carnot engine consists of two sets of alternating \textit{adiabatic} strokes and \textit{isothermal} strokes. The quantum analogue of the Carnot engine consists of a working fluid, which can be a particle in a box \cite{bender2000quantum}, qubits \cite{linden2010small,geva1992quantum}, multiple level atoms \cite{quan2007quantum} or harmonic oscillators \cite{kosloff1984quantum,lin2003performance}. We emphasize that for all such engines, the efficiency of the engine is strictly bounded by the Carnot efficiency \cite{gardas2015thermodynamic}. For the engine consisting of non-interacting qubits considered in \cite{geva1992quantum}, the Carnot cycle consists of 
\begin{enumerate}
\item \textit{Adiabatic Expansion}: An expansion wherein the spin is uncoupled from any heat baths and its frequency is changed from $\omega_1$ to $\omega_2>\omega_1$ adiabatically. Work is done by the spin in this step due to a change in the internal energy. Since the spin is uncoupled, its von Neumann entropy is conserved in this expansion stroke.
\item \textit{Cold Isotherm}: The spin is coupled to a cold bath at inverse temperature $\beta_C$. This transfers heat from the engine to the cold bath.
\item \textit{Adiabatic Compression}: A compression stroke where the spin is uncoupled from all heat baths and its frequency is changed from $\omega_3$ to $\omega_4$ adiabatically. Work is done on the medium in this step.
\item \textit{Hot Isotherm}: The spin is coupled to a hot bath at inverse temperature $\beta_H<\beta_C$. This transfers heat to the engine from the hot bath.
\end{enumerate}

\begin{figure}[t]
\centering
	\includegraphics[width=0.35\textwidth]{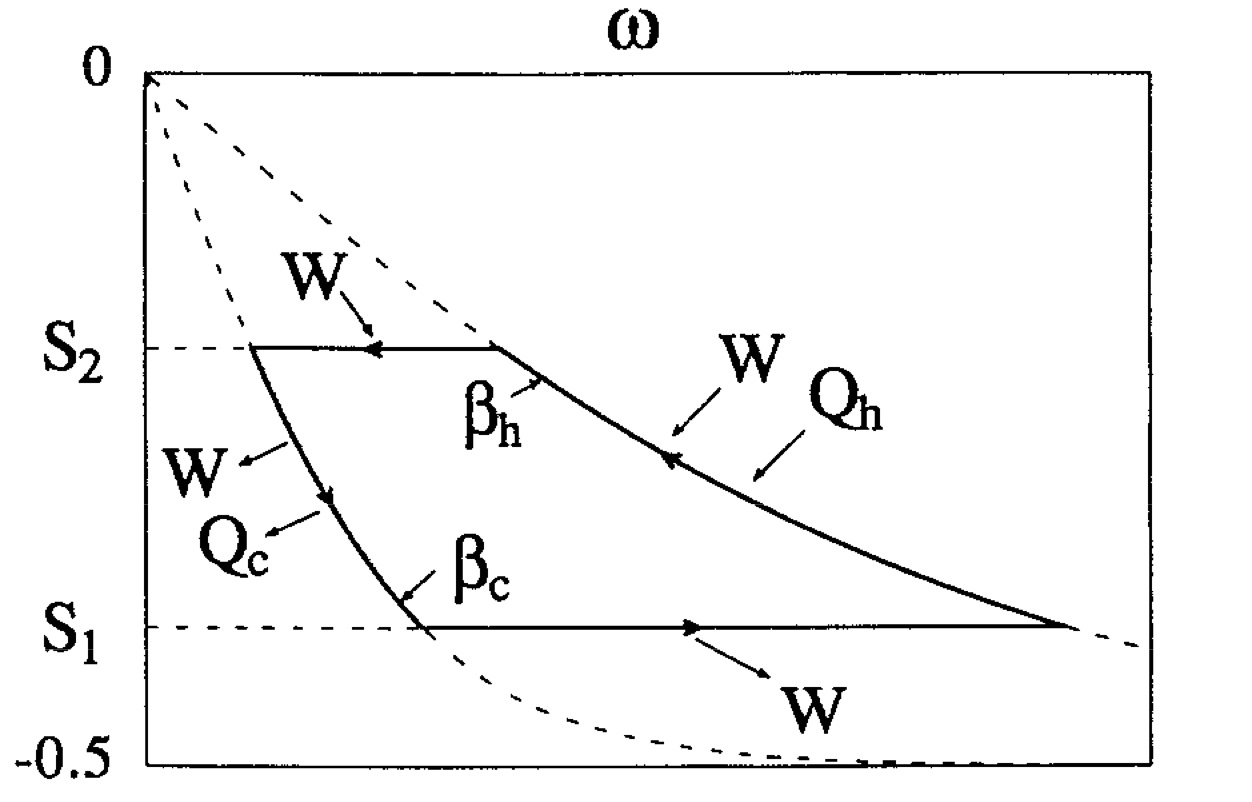} \quad \quad 
	\includegraphics[width=0.35\textwidth]{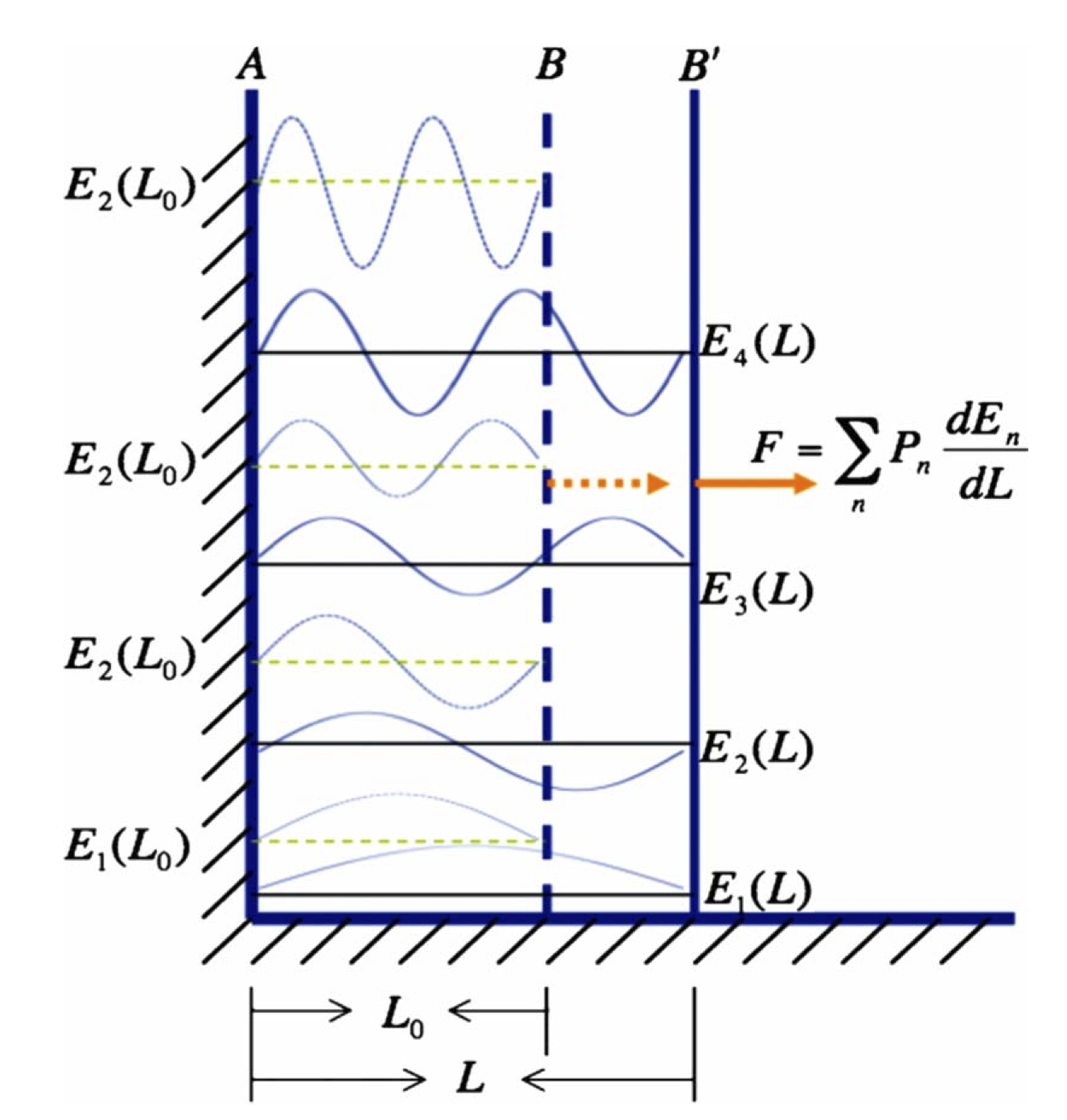} 
	\caption{\label{fig:GQ_Comb} The left figure is a reversible Carnot cycle, operating in the limit $\dot{\omega}\rightarrow0$, depicted in the space of the normalised magnetic field $\omega$ and the magnetisation $\text{S}$. The horizontal lines represent adiabats wherein the engine is uncoupled from the heat baths and the magnetic field is changed between two values. The lines connecting the two horizontal strokes constitute changing the magnetisation by changing $\omega$ while the qubits are connected to a heat bath at constant temperature. We note that in the figure, $\text{S}_{1,2}<0$ represent two values of the magnetisation, with $\text{S}_1<\text{S}_2$.\borrowedfig{(Figure taken from {\it Jour. Chem. Phys. } {\bf 96,} 3054 (1992).)} The right figure is a particle in a box with engine strokes being defined in terms of expansion and contraction of the box. This defines a quantum isobaric process, by defining force. This definition is used to define a Diesel cycle in the text. We note that in the figure, force is denoted by $F$, while its denoted by $\mathbb{F}$ in the text. \borrowedfig{(Figure taken from {\it Phys. Rev. E} {\bf 79,} 041129 (2009).)}}
\end{figure}

The inverse temperature $\beta'$ of the thermal state of this working fluid is given at any point $(\omega,\text{S})$ on the cycle, by the magnetisation relation $\text{S}=-\tanh(\beta'\omega/2)/2$. The dynamics is described by a Lindblad equation, which will then be used to derive the behaviour of heat currents. The engine cycle is described in Fig.~\ref{fig:GQ_Comb}. The Hamiltonian describing the dynamics is given by $H(t)=\omega(t)\sigma_3/2$. The Heisenberg equation of motion is written in terms of $\mathcal{L}_{D}(\sigma_3)$, the Lindblad operators describing coupling to the baths at inverse temperatures $\beta_H$ and $\beta_C$ respectively. This equation can be used to calculate the expectation value of the Hamiltonian $\langle H(t)\rangle$, which leads to
\begin{align}\label{first_first}
\frac{\d \langle H(t)\rangle}{\d t}=\frac{1}{2}\left(\frac{\d\omega}{\d t}\langle \sigma_3\rangle + \omega \langle \mathcal{L}_{D}(\sigma_3) \rangle \right)=\frac{1}{2}\left(\frac{\d\omega}{\d t}\langle \sigma_3\rangle+\omega\frac{\d\langle \sigma_3\rangle}{\d t}\right).
\end{align}
As described in section \ref{sec:qt}, the definition of work $\langle\delta W\rangle=\langle \sigma_{3}\rangle\delta\omega/2$ and heat $\langle\delta Q\rangle=\omega\delta\langle \sigma_{3}\rangle/2$ emerge naturally from the above discussion, and are identified with the time derivative of the first law. For \sai{this} quantum Carnot engine, the maximal efficiency which is that of the reversible engine  \cite{geva1992quantum} \sai{is given by the standard formula namely}
\begin{align}
\eta_{\rm Carnot}=1-\frac{T_{\rm C}}{T_{\rm H}}.
\end{align}

\subsection{Otto Engine}
As a counterpoint to the Lindblad study of Carnot engines implemented via qubits, let us consider the harmonic oscillator implementation of Otto engines \cite{scully2002quantum,abah2012single,deng2013boosting,del2014more,zheng2014work} whose time dependent frequency of the harmonic oscillator working fluid is $\omega^{\rm HO}(t)$.  The initial state of the oscillator is a thermal state at inverse temperature $\beta_{\rm C}$, and the oscillator frequency is $\omega^{\rm HO}_1$. The four strokes of the Otto cycle are given by:
\begin{enumerate}
\item \textit{Adiabatic Compression}: An compression, where the medium is uncoupled from all heat baths and its frequency is changed from $\omega^{\rm HO}_1$ to $\omega^{\rm HO}_2>\omega^{\rm HO}_1$. Since the oscillator is uncoupled, its von Neumann entropy is conserved during this stroke, though it no longer is a thermal state.
\item \textit{Hot Isochore}: At frequency $\omega^{\rm HO}_2$, the harmonic oscillator working fluid is coupled to a bath whose inverse temperature is $\beta_{\rm H}$, and is allowed to relax to the new thermal state.
\item \textit{Adiabatic Expansion}: In this stroke, the medium is uncoupled from any heat baths and its frequency is changed from $\omega^{\rm HO}_2$ to $\omega^{\rm HO}_1$. 
\item \textit{Cold Isochore}: At frequency $\omega^{\rm HO}_1$, the harmonic oscillator working fluid is coupled to a bath whose inverse temperature is $\beta_{\rm C}$ and it is allowed to relax back to its initial thermal state.
\end{enumerate}

Since the two strokes where the frequencies change, work is exchanged, whereas the two thermalising strokes involve heat exchanges, the efficiency is easily calculated to be
\begin{align}
\eta_{\rm Otto}=-\frac{\langle \delta W\rangle_1+\langle \delta W\rangle_3}{\langle \delta Q\rangle_2}.
\end{align}

\begin{figure}[t]
\centering
	\includegraphics[width=0.3\textwidth]{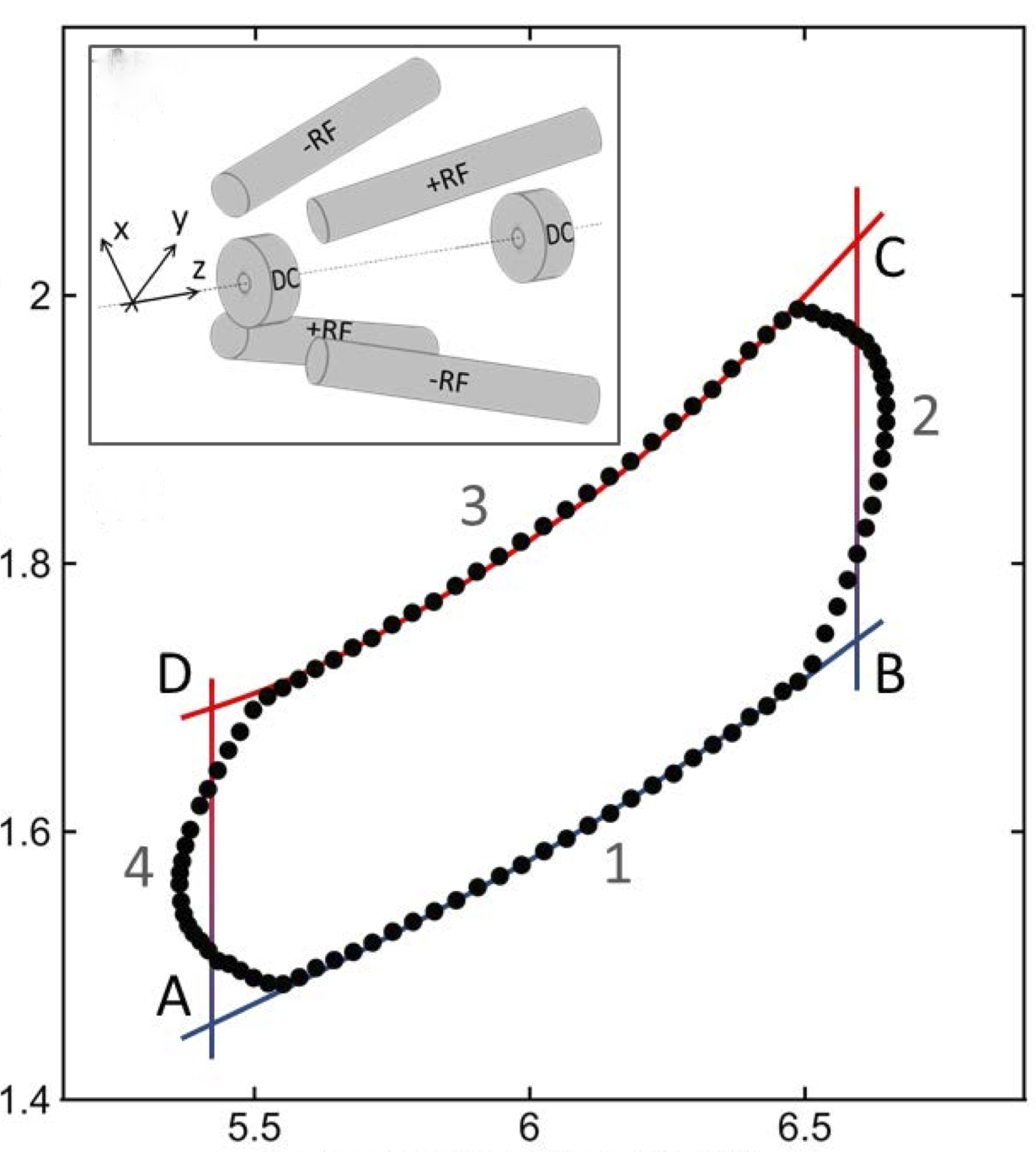} \quad \quad  \quad \quad 
	\includegraphics[width=0.20\textwidth]{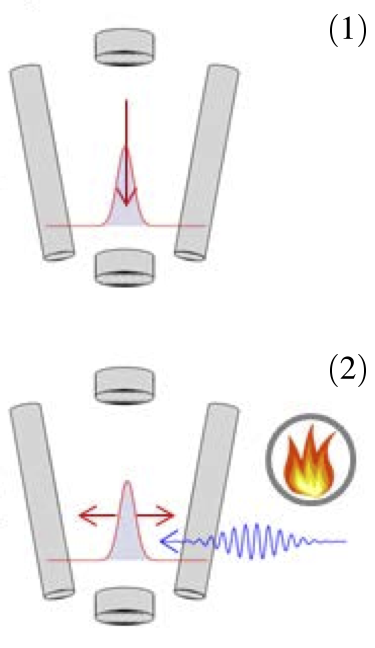} \quad  
	\includegraphics[width=0.20\textwidth]{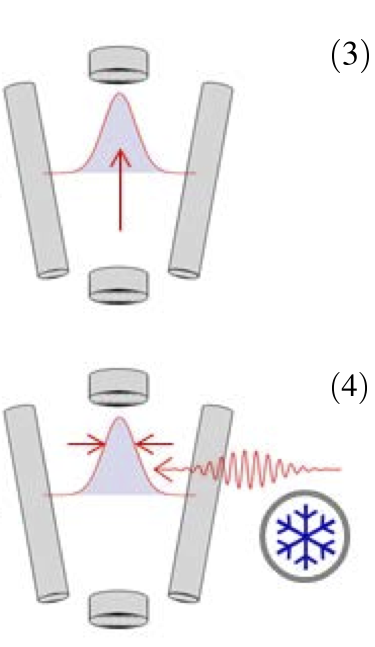}
	\caption{\label{fig:QTM_EXPT} Experimental proposal of an Otto engine consisting of a single trapped ion, currently being built by K. Singer's group \cite{NJP_QT}. On the left, the energy-frequency diagram corresponding to the Otto cycle implemented on the radial degree of freedom of the ion. The inset on the left hand side is the geometry of the Paul trap. On the right hand side, a representation of the four strokes, namely (1) adiabatic compression, (2) hot isochore, (3) adiabatic expansion and (4) cold isochore, see text for details. \borrowedfig{(Figure taken from {\it Phys. Rev. Lett. } {\bf 112,} 030602 (2014).)}}
\end{figure}

The experimental implementation of such a quantum Otto engine was considered in \cite{abah2012single,rossnagel2014nanoscale} and is presented in Fig.~\ref{fig:QTM_EXPT}. The Paul or quadrupole trap uses rapidly oscillating electromagnetic fields to confine ions using an effectively repulsive field. The ion, trapped in the modified trap, presented in Fig.~\ref{fig:QTM_EXPT} is initially cooled in all spatial directions. The engine is coupled to a hot and cold reservoirs composed of blue and red detuned laser beams. The tapered design of the trap translates to an axial force that the ion experiences. A change in the temperature of the radial state of the ion, and hence the width of its spatial distribution, leads to a modification of the axial component of the repelling force. Thus heating and cooling the ion moves it back and forth along the trap axis, as induced in the right hand side of the figure. The frequency of the oscillator is controlled by the trap parameters. Energy is stored in the axial mode and can be transferred to other systems and used.

\sai{Since we have considered an Otto engine operating between two thermal baths, we should not be surprised that the standard formula for efficiency still applies. Making one of the components of the engine genuinely non-thermal (and quantum) makes the analysis more interesting. In \cite{rossnagel2014nanoscale}, the authors show that the standard Carnot efficiency can be overcome by squeezing the thermal baths. In \cite{abah2014efficiency}, the authors derive a ``generalised" Carnot efficiency that correctly accounts for the first and second law. Using this analysis, they demonstrate that though the engine efficiency exceeds the standard Carnot formula, the ``generalized" efficiency is not exceeded, in keeping with the laws of thermodynamics \cite{niedenzu2015efficiency}. We note that the cost of squeezing the bath has not been accounted for, similar to how the cost of preparing a cold bath is unaccounted for in classical thermodynamics (it is assumed that the baths are free resources).}

\subsection{Diesel Engine}
Finally, we discuss Diesel cycles designed with a particle in a box as the working fluid \cite{quan2009quantum}. The classical Diesel engine is composed of isobaric strokes, where the pressure is held constant. We will discuss how the notion of ``pressure" is generalised to the quantum setting and how this defines a ``quantum isobaric" stroke. To define pressure, we begin by defining force. After defining average work and heat as before, generalised forces can be defined by analogy with classical thermodynamics using the relation
\begin{align}
Y_n=\frac{\langle\delta W\rangle}{\delta y_n},
\end{align}
where ${y_n,Y_n}$ form the $n$-th conjugate pair in the definition of work $\langle \delta W\rangle=\sum_nY_n\delta y_n$. \sai{Since work is defined for such continuous machines as $\langle\delta W\rangle=\tr [ \rho \, \delta H]$ \cite{alicki1979quantum}}, for eigenstates with energies $E_n$ distributed according to a probability distribution $P_n$ \sai{the force $ \mathbb{F}$ is given by}
\begin{align}
 \mathbb{F}=\displaystyle\sum_nP_n\frac{\delta E_n}{\delta L}.
\end{align}

If the system is in equilibrium with a heat bath at inverse temperature $\beta$, with the corresponding free energy being $F=-\log(Z)/\beta$, then the force is given by the usual formula $\mathbb{F}=-[\d F/\d L]_{\beta}$, evaluated at constant temperature $\beta^{-1}$. This force is calculated to be $ \mathbb{F}=(\beta L)^{-1}$. Hence to execute an ``isobaric process", wherein the pressure is held constant across a stroke, the temperature must vary as $\beta=( \mathbb{F}L)^{-1}$. This can be employed to construct a Diesel cycle, presented in Fig.~\ref{fig:GQ_Comb}, and whose four strokes are given by
\begin{enumerate}
\item \textit{Isobaric Expansion}: The expansion of the walls of the particle in a box happen at constant pressure. The width of the walls goes from $L_2$ to $L_3>L_1$.
\item \textit{Adiabatic Expansion}: An adiabatic expansion expansion process, wherein the length goes from $L_3$ to $L_1>L_3$. Entropy is conserved in this stroke.
\item \textit{Isochoric Compression}: The compression happens at constant volume $L_1$ where the force on the box is reduced.
\item \textit{Adiabatic Compression}: The compression process takes the box from volume $L_1$ to $L_2<L_1$. This is done by isolating the quantum system, conserving the entropy in the process.
\end{enumerate}
Like before, the efficiency is calculated by considering the ratio of the net work done by the system to the heat into the system. A straightforward calculation of this efficiency yields
\begin{align}
\eta_{\rm Diesel}=1-\frac{1}{3}(r^2_{\rm E}+r_{\rm E}r_{\rm C}+r^2_{\rm C}).
\end{align}
Here $r_{\rm E}=L_3/L_1$ is the expansion ratio and $r_{\rm C}=L_2/L_1$ is the compression ratio.

\subsection{Quantumness of Engines}

\sai{Since the design of quantum engines is often analogous to their classical counterparts, a central question is: what are genuinely quantum ingredients for engines?} There are two important points to consider here: The first point is that when thermal machines operate between two heat baths with well defined temperatures, the efficiency of such a machine is also limited by the classical efficiencies, i.e. the Carnot and Curzon-Ahlborn efficiencies. \sai{The Carnot efficiency for instance, is derived independent of the details of the working fluid and depends only on the laws of thermodynamics. It is expected that these bounds must also hold in the quantum setting.}

Once the thermal machines are committed to operate between two temperatures and in equilibrium, not much can be done to change the efficiency of these machines (see \cite{niedenzu2015performance} for a recent example). Hence the deviation from classical performance of these thermal machines is only seen when either the baths are made non-thermal or the working fluid of the thermal machine is not allowed to equilibrate, operating in a non-equilibrium setting. To characterize such non-equilibrium thermal machines, several authors have studied the role of quantum correlations \cite{gallego2014thermal} and the role they play in work extraction. 

\sai{Using non-thermal heat baths \cite{scully2003extracting,rossnagel2014nanoscale} has been another strategy to see the effect of quantum states on thermal machines. An example of such a study focuses on Otto engines operated with squeezing thermal baths we discussed in the previous subsection. In all such examples, non-classical resources are employed to improve engine performance, from power generation \cite{del2014more,binder2015quantacell} to efficiency \cite{abah2012single,rossnagel2014nanoscale,abah2014efficiency}.} Finally, we point out that time has an important role to play in engine performance, since it relates to power. As detailed in \ref{sub:CTTM}, both coherence and time play important roles in the design of quantum machines \cite{geva1992classical,PL15,Huber15}.  As an example, consider that the evolution of an initial state to a final state cannot happen faster than a fundamental bound that depends on the Hamiltonian and states involved. The minimum time associated with this bound is often called the quantum speed limit \cite{mandelstam1945uncertainty}. The quantum speed limit places a limit on the power of quantum engines, and has been used to improve engine performance \cite{binder2015quantacell}. Another approach to time in quantum thermodynamics employs so-called adiabatic shortcuts to improve engine performance in both the classical \cite{deng2013boosting} and quantum regime \cite{del2014more}. 

  An adiabatic quantum system that starts in an eigenstate of the Hamiltonian remains in the instantaneous eigenstate of the Hamiltonian when the parameters of the Hamiltonian are swept slowly enough (distinguish this from the thermodynamic definition of adiabaticity, which relates to having zero heat exchange). This typically slow dynamics can be sped up both in the classical and quantum contexts by the addition of external control fields that are bound to be transitionless. Finally we note that more foundational aspects of quantum mechanics, such as non-commutativity of operators, have been shown to have a detrimental effect on engine performance. This is known as quantum friction and is also a genuinely quantum effect that impacts engine performance \cite{kosloff2002discrete}. \sai{Besides issues such as non-thermal baths and non-commutativity}, we refer the reader to \cite{de2014quantum} for an information theoretic perspective on engine design.
  
\sai{To answer the question about the quantumness of engines, recent studies have focussed on trying to study alternative non-quantum models (to compare and contrast with quantum engines) or produce genuine quantum effects to demonstrate the quantumness of engines}. On the difference between quantum mechanics and stochastic formulation of thermodynamics, in \cite{uzdin2015quantum} the authors study  quantumness of engines and show that there is an equivalence between different types of engines, namely two-stroke, four-stroke and continuous engines. Furthermore, the authors define and discuss quantum signatures in thermal machines by comparing these machines with equivalent stochastic machines, and they demonstrate that quantum engines operating with coherence can in general output more power than their stochastic counterparts. Finally, we note an example of a quantum thermal machine \cite{brask2015autonomous} operating in the continuous time regime \sai{that produces entanglement.} The working fluid \sai{considered by the authors consists of} two qubits, which are coupled to two different baths and to each other. The steady state of this thermal machine far from thermal equilibrium (in the presence of a heat current) exhibits the interesting property of creating bipartite entanglement in the working fluid. \sai{Such an engine can be said to have a ``genuinely quantum" output.}

\subsection{Quantum Refrigerators}
Refrigerators are engines operating in a regime where the heat flow is reversed. Like engines, the role that quantised energy states, coherence and correlations play in the operation of a quantum refrigerator have been fields of extensive study
\cite{correa2013performance,correa2014quantum,kosloff2014quantum,ticozzi2014quantum,uzdin2014multilevel,brunner2014entanglement,Brask15}.

\begin{figure}[t]
\centering
	\includegraphics[width=0.45\textwidth]{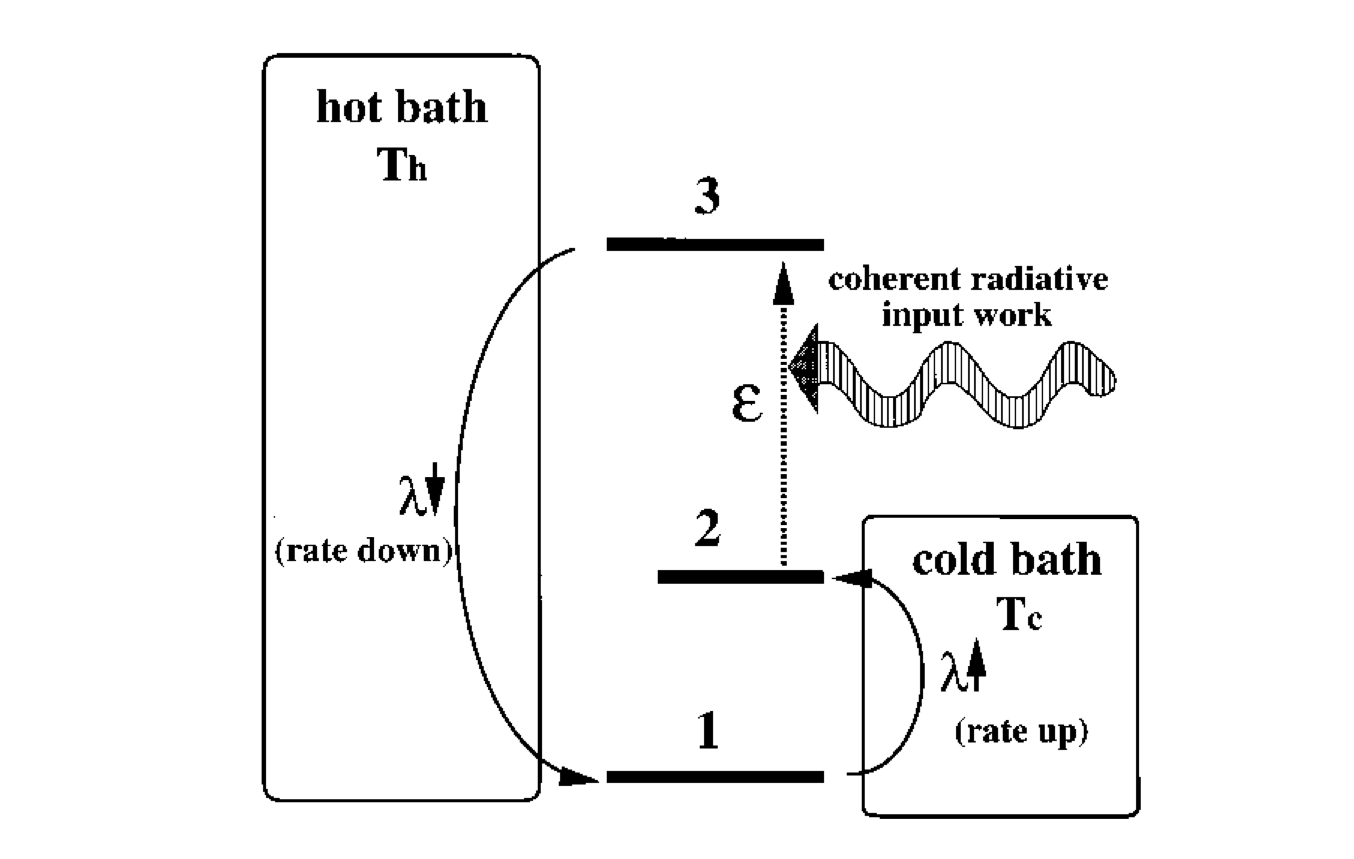} \quad \quad \quad 
	\includegraphics[width=0.42\textwidth]{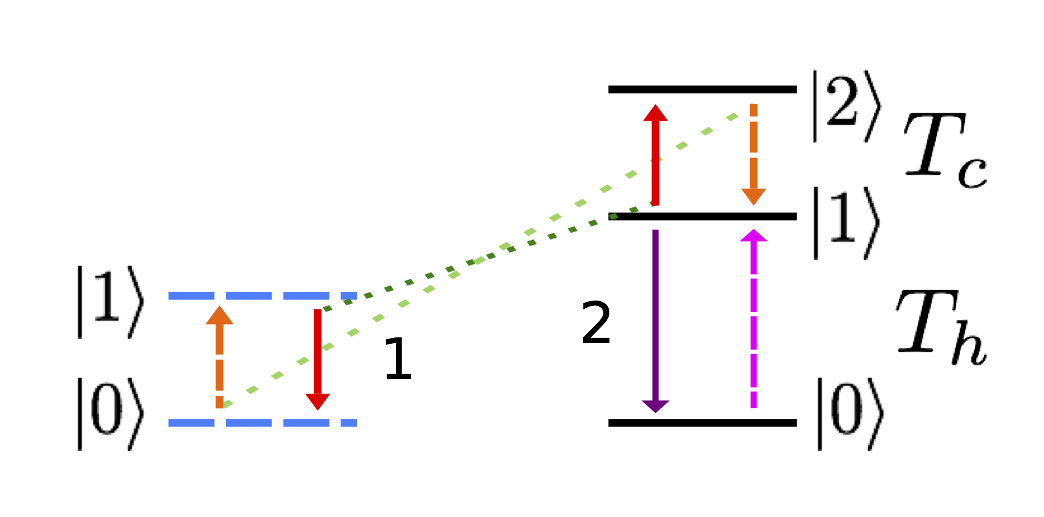} 
	\caption{\label{fig:GP_Comb} Two designs of refrigerators, the details of which are discussed in the text. The left figure consists of a three-level atom coupled to two baths and interacting with a field.  \borrowedfig{(Figure taken from {\it Jour. Appl. Phys.} {\bf 87,} 8093 (2000).)}. The right figure consists of the smallest refrigerator, composed of a qubit and a qutrit. Here, cooling of the qubit is achieved through joint transformations between the system and the reservoir, wherein the system's entropy is shunted to the reservoir. \borrowedfig{(Figure taken from {\it Phys. Rev. Lett. } {\bf 105,} 130401 (2010).)} }
\end{figure}

We present two examples of refrigerator cycles considered in the literature. The first of these \cite{kosloff2000quantum} is built on a three level atom, coupled to two baths, as shown in Figure \ref{fig:GP_Comb}. The refrigerator is driven by coherent radiation. This induces transitions between level 2 and \sai{level} 3. The \sai{population in} level 3 then relaxes to level 1 by rejecting heat to the hot bath at temperature \sai{$T_{\rm H}$}. The system then transitions from level 1 to level 2 by absorbing energy from a cold bath. Like before, the dynamics can be written in terms of the Heisenberg equation for an observable $A$.  Let \sai{$\mathcal{L}_{\rm{C,H}}$} refer to the Lindblad operators corresponding to the cold and hot baths respectively. Since this is a thermal machine operated continuously in time, the steady state energy transport is the measure of heat transported by the machine. Assuming the Hamiltonian $H=H_0+V(t)$, the energy transport can be written as 
\begin{align}
\frac{\d\langle H\rangle}{\d t}=\left\langle\frac{\partial V(t)}{\partial t}\right\rangle +\langle \mathcal{L}_{\rm H}(H)\rangle+\langle \mathcal{L}_{\rm C}(H)\rangle.
\end{align}
Using this, the authors in \cite{kosloff2000quantum} investigated the refrigerator efficiency. This efficiency is expressed in terms of the coefficient of performance (COP), which is defined as the ratio of the cooling energy to the work input into the system. For two thermal baths at temperatures \sai{$T_{\rm C}$ and $T_{\rm H}$} respectively, the universal reversible limit is the Carnot COP, given by
\begin{align}
\text{COP}=\sai{\frac{T_{\rm C}}{T_{\rm H}-T_{\rm C}}}.
\end{align}
COP can be larger than one, and in \cite{palao2001quantum}, the authors show the relationship of \sai{the COP to the cooling rate}, which is simply the rate at which heat is transported \sai{from} the system.

The other example of quantum refrigerators takes a more quantum information theoretic point of view. In \cite{linden2010small}, the authors were inspired by algorithmic cooling and studied the smallest refrigerators possible. One of the models consists of a qubit coupled to a qutrit, \sai{see} Fig. \ref{fig:GP_Comb}. The intuition for cooling the qubit comes from a computational model of entropy reduction.  This procedure, called algorithmic cooling \cite{schulman1999molecular}, is a procedure to cool quantum systems which uses ideas of entropy shunting to ``move" entropy around a large system in a way that lowers the entropy of a subsystem. Consider $n$ copies of a quantum system which is not in its maximal entropy state. Joint unitary operations on the $n$ copies can allow for the distillation of a small number $m$ of systems which are colder (in this context, of lower entropy) than before, whilst leaving the remaining $(n-m)$ systems in a state that is hotter than before (this is required by unitarity). This is the intuition behind cooling of the qubit in Fig.~\ref{fig:GP_Comb}. The refrigerator in \cite{linden2010small} is the smallest (in Hilbert space dimensionality) refrigerator possible. See \cite{correa2013performance} for a discussion of the relationship of power to COP and \cite{correa2013performance} for a discussion of correlations in the design of quantum absorption refrigerators. Finally, we note that there is a connection between models of cooling based on study of heat flows of continuous models, such as sideband cooling, and quantum information theoretic models of cooling based on control theory \cite{wang2011ultraefficient,wang2013absolute,horowitz2015energy}.

\subsection{Coherence and Time in Thermal Machines}\label{sub:CTTM}
In this subsection, we want to briefly discuss the role coherence and time play in quantum thermal machines. The fundamental starting point of finite time classical \cite{andresen2011current} and quantum thermodynamics is the fact that the most efficient macroscopic processes are also the ones that are quasistatic and hence slow. Hence they are impractical from the standpoint of power generation. The role of quantum coherences was illustrated by a series of studies involving commutativity \sai{of parts of the Hamiltonian, as explained below}. \sai{Consider a quantum Otto engine, whose energy balance equation is given in terms of the Hamiltonian part $H_{\rm ext}+H_{\rm int}$ and dissipative Lindblad operator $\mathcal{L}$ \cite{feldmann2003quantum}. There are two sources of heat that were studied in this model, and they are related to coherence and time. The first of these sources of heat are attributed to energy transfer from the hot bath to the cold bath via the system. The second source of heat comes from the finite time driving of the adiabatic strokes of the engine. Such a finite time driving means that the system causes irreversible heating which are understood as follows. Suppose the initial state of the system is an eigenstate of $H_{\rm{ext}}$. If $[H_{\rm ext},H_{\rm int}]\neq0$, then the quantum state, even in the absence of Lindblad terms, does not ``follow" the instantaneous eigenstate of the external Hamiltonian. The precession of the quantum state of the system about this instantaneous eigenstate, that is induced by the non-commutativity of the external and internal Hamiltonians is a source of a friction like heat \cite{kosloff2002discrete}.}

\begin{figure}[t]
\centering
	\includegraphics[width=0.75\textwidth]{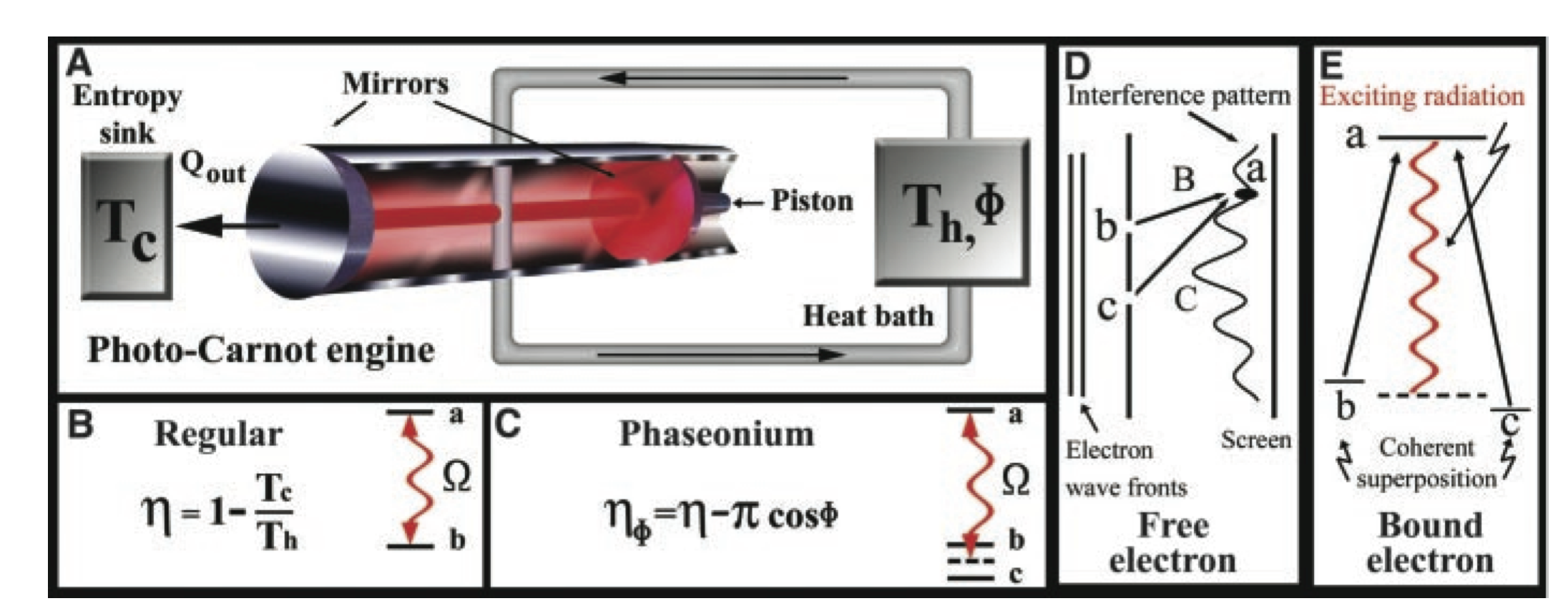} 
	\caption{\label{fig:SCULLY} An optical Carnot engine, discussed in the text. The engine is driven by the radiation in the cavity being in equilibrium with atoms with two separate temperatures during the hot and cold part of the Carnot cycle. The phased three level atoms have an off-diagonal coherence term $\Phi$ that is used to improve the Carnot efficiency.\borrowedfig{(Figure taken from {\it Science} {\bf 299,} 862 (2003).)}}
\end{figure}

\sai{Finally, we present another model of a non-thermal bath consisting of three-level atoms.} In \cite{scully2003extracting}, the authors extract work from coherences by using a technique similar to lasing without inversion. The photon Carnot engine proposed by the authors to study the role of coherences follows by considering a cavity with a single mode radiation field enclosed in it. The radiation interacts with phased three level atoms at two temperatures during the isothermal parts of the Carnot engine cycle. The three level atoms are phased by making them interact with an external microwave source. The efficiency of this engine was calculated to be
\begin{align}
\eta=\eta_{\rm Carnot}-\pi\cos(\Phi),
\end{align}
where the phased three level atoms have an off-diagonal coherence term $\Phi$ that is used to improve the Carnot efficiency. This engine is depicted in Fig.~\ref{fig:SCULLY}. Thus by employing the coherence dynamics of three level atoms, the authors demonstrate enhancement in work output. \sai{We end this section by noting that quantum features are not necessarily beneficial to a QTM's performance. For example, a study of entanglement dynamics in three-qubit refrigerators \cite{brunner2014entanglement} demonstrates that there is only very little entanglement found when the machine is operated near the Carnot limit.} 

\subsection{Correlations, Work Extraction \& Power}\label{sub:CWE}

Finally, let us discuss aspects of work extraction common to both cyclical and non-cyclical thermal machines. A non-cyclical example of the use of quantum correlations to improve machine performance involves quantum batteries. These are quantum work storage devices and their key issues relate to their capacity and the speed at which work can be deposited in them. This question of work deposition can be studied in two steps. Firstly, it is desirable to understand the limits on the work extractable from a quantum system under unitary transformations, V. Here, consider states to be written in an ordered eigenbasis as $\rho=\sum_k r^{\uparrow}_{k}\vert r^{\uparrow}_{k}\rangle\langle r^{\uparrow}_{k}\vert$ and $H=\sum_k \epsilon^{\downarrow}_k \vert \epsilon^{\downarrow}_{k}\rangle\langle \epsilon^{\downarrow}_{k}\vert$. Note that $\vert r^{\uparrow}_{k}\rangle$, the eigenstates of $\rho$, are a-priori unrelated to $\vert \epsilon^{\uparrow}_{k}\rangle$, the eigenstates of $H$. The arrows show the ordering of eigenvalues, namely $r^{\uparrow}_{0}\leq r^{\uparrow}_{1}\ldots$ and likewise for the Hamiltonian. The optimal unitary that extracts the most work from $\rho$ is given by the unitary that maps $\rho$ to the state $\pi=\sum r^{\uparrow}_{k}\vert \epsilon^{\downarrow}_{k}\rangle\langle \epsilon^{\downarrow}_{k}\vert$ \cite{PW,LENARD, allahverdyan2004maximal,AG13}.
Such states, where the eigenvalues are ordered inversely with respect to the eigenvectors of the Hamiltonian, are known as ``passive states" \cite{PW,LENARD} and the corresponding maximum work extracted from the unitary transformation, 
\begin{align}
	\< W^{\max}_{\rm ext} \> =\tr [H (\rho - \pi)].
\end{align}
is called ``ergotropy". Note, that the thermal states $\tau_{\beta} = e^{-\beta H}/ \tr [e^{-\beta H}]$ for any inverse temperature $\beta$ are passive. Ground states of a given Hamiltonian, for instance, are examples of states passive with respect to the Hamiltonian and correspond to $\beta=\infty$. \add{Thermal states maximize the entropy for a given energy and minimize the energy for a given entropy. Passive states which maximize the energy for a given entropy were sought in \cite{perarnau2015most}, and are useful in bounding the cost of thermodynamic processes. Passive states are important to understand the functioning of quantum engines, see \cite{gelbwaser2013work} for example.} 

The role entangling operations play in extracting work using unitary transformations was explored within the simultaneous extraction of work from $n$ quantum batteries \cite{PhysRevE.87.042123}. They considered $n$ copies of a given state $\sigma$ and a Hamiltonian $H$ for each system. As discussed before, $\sigma$ and $H$ imply that there exists a passive state $\pi$. Two strategies for extracting work exist, the first of these being to simply process each quantum system separately. The second strategy is to do joint entangling operations, which transform the initial state $\sigma^{\otimes n}$ to a final state which has the same entropy as $n$ copies of $\sigma$. \sai{The authors considered a Gibbs state $\tau_{\beta}$ where the temperature was fixed by matching the entropy of this state with $\sigma$. They considered such a state since this state provided a upper bound on the amount of work that could be extracted by transforming $\sigma^{\otimes n}$ into a passive state.} Since the final state $\tau_{\beta}^{\otimes n}$ has the same entropy as $n$ copies of the initial state, a transformation between $\sigma^{\otimes n}$ and the neighbourhood of $\tau_{\beta}^{\otimes n}$ would be the desired transformation that  extracts maximum work. Such a transformation was shown to be entangling, using typicality arguments \cite{wilde2013quantum}. However, the authors in \cite{PhysRevLett.111.240401} pointed out that there is always a protocol to extract all the work from non-passive states by using non-entangling operations, although in comparison to entangling unitaries more non-entangling operations are needed. This suggests a relationship between power and entanglement in work extraction, an assertion demonstrated for quantum batteries in \cite{binder2015quantacell}. A different role can be played by correlations in storing and extracting work, \sai{as discussed in Sec.~2}. This role relates to storing work in correlations. It has been shown by several authors \cite{brunner2014entanglement,PL15,Huber15} that measures of correlations like quantum mutual information and entanglement can affect the performance of thermal machines.

A final noteworthy point in the consideration of time explicitly in the context of quantum thermal machines relates to maximum power engines. Firstly, since the optimal performance of engines corresponds to quasi static transformations, the power is always negligible and the time of operation of one cycle is infinite (or simply much longer compared to the internal dynamical timescales of the quantum systems). To remedy this, both classically and quantum mechanically, engines which optimise power have been considered. These studies find the efficiency at maximum power, see Eq.~(\ref{eq:CA}), first derived in the classical context by Curzon and Ahlborn \cite{curzon1975efficiency} but also valid in the quantum case \cite{kosloff1984quantum}. See \cite{del2014more} for a discussion of shortcuts to adiabaticity and their role in optimising power in Otto engines and \cite{correa2014multistage} for a calculation of efficiency at maximum power of an absorption heat pump, which also proved the weak dependence of efficiency and power with dimensionality. In \cite{rezek2006irreversible,abah2012single}, the authors consider the power of Otto cycles in various regimes, and derive the steady-state efficiency
\begin{align}
\eta_{\rm SS}=\frac{1-\sqrt{\beta_{\rm H}/\beta_{\rm C}}}{2+\sqrt{\beta_{\rm H}/\beta_{\rm C}}}.
\end{align}
\subsection{Relationship to Laws of Thermodynamics}
We end the discussion of QTMs with a small report on the role that the laws of thermodynamics, discussed in section \ref{sec:qt}, play in constraining design. As was noted in the context of engines, the first law emerges naturally as a partitioning of the energy of the system in terms of heat and work. See \cite{binder2015quantum} for a partitioning of internal energy change for CPTP processes. In the context of QTMs, for instance, this can be seen in Eq.~(\ref{first_first}). The second law manifests itself usually by inspecting the entropy production of the universe \cite{kosloff2000quantum}. This is given for a typical \sai{continuous regime} QTM, with a system and two baths as
\begin{align}
\frac{\d S}{\d t}=-\frac{\langle \dot{Q}_{\rm H}\rangle}{T_{\rm H}}-\frac{\langle \dot{Q}_{\rm C}\rangle}{T_{\rm C}}.
\end{align}
Analysis of the dynamics shows that the second law enforces the rule that the net entropy production, given by the equation above, is non-negative. See \cite{geva1996quantum,correa2013performance} to study the relationship between weak coupling and the second law. Information theoretic approaches often use monotonicity condition of relative entropy to track entropy production, see section \ref{sec:qdft}. Finally, we note the work in \cite{uzdin2014multilevel} relating to the second law and \cite{campisi2015nonequilibrium} for heat engine fluctuation relation and experimental proposal, all in the context of SWAP engines. In \cite{levy2014local}, the authors discuss how in a network of quantum systems coupled to reservoirs at different temperatures, the second law is violated locally, though its always valid globally, in line with intuition (see references in \cite{vinjanampathy2014second} for a discussion on the violation of entropic inequalities, see \cite{martinez2013dynamics} for a discussion on the role of non-linear couplings in refrigerator design). \sai{The role of the second law, in all these examples, tends to be to modify the existing figures of merit, like efficiency \cite{rossnagel2014nanoscale} and hence show a path towards understanding nanoscale QTMs.}

The third law states that the entropy of any quantum system goes to a constant as temperature approaches zero. This constant is commonly assumed to be zero. This means, that in the context of the equation above, the heat current corresponding to the cold bath, $\langle \dot{Q}_{\rm C}\rangle\propto T^{\alpha+1}_{\rm C}$ as $T_{\rm C}\rightarrow 0$ with $\alpha>0$. \sai{Since we are discussing the limit of the cold bath temperature going to zero, we expect that the third law will inform the operation of quantum refrigerators and heat engines, where the cold bath temperature is quite low. Furthermore, the third law affects the ability for us to cool a quantum system as the system approaches absolute zero. This is because there is a tradeoff between the entropy generated by the system coupling to an environment that is used to cool the system and the fact that approaching absolute zero means that the system becomes a pure state (assuming the ground state is unique). This tradeoff practically limits COP.} This was studied in \cite{kosloff2000quantum}, see also \cite{kosloff2014quantum} for a discussion on another formulation of the third law and \cite{kosloff2013quantum} for a recent review of the laws of thermodynamics in the quantum regime. In \cite{kolavr2012quantum}, the authors discuss periodically driven quantum systems and investigate the laws of thermodynamics for a model of quantum refrigerator, see \cite{alicki2012periodically,gelbwaser2013minimal,alicki2015non}. Finally, in \cite{levy2012quantum}, the authors study the third law in the context of refrigerators and show that the rate of cooling is determined entirely by the cold reservoir and its interaction to the system, and is insensitive to underlying the particle statistics.
\section{Discussion and Open Questions} \label{sec:discussion}

We have presented an overview of a selection of current approaches to quantum thermodynamics pursued with various techniques and interpreted from different perspectives. Substantial insight has been gained from these advances and their combination, and a unified language is starting to emerge. This difference in perspectives has also meant that there are ideas within quantum thermodynamics where consensus is yet to be established. 

One example of such disagreement is the definition of work in the quantum regime. Various notions of work have been introduced in the field, including the average work defined for an ensemble of experimental runs in Eq.~(\ref{eq:Q+W}), classical and quantum fluctuating work defined for a single experimental run in Eq.~(\ref{eq:zeroQ}) and Eq.~(\ref{eq:qW}), optimal single shot work given in Eq.~(\ref{eq:singleshotwork}) and optimal thermodynamic resource theory work given in Eq.~(\ref{eq:ho}). It is reassuring to see that the latter two, quite separate single shot approaches, result in the same optimal work value. Despite advances in unifying these work concepts our understanding of work in the quantum regime remains patchy. For example, resource theory work Eq.~(\ref{eq:ho}) is a work associated with an \emph{optimised thermal operations process} - to move a work storage system almost deterministically as high as possible - while the fluctuating work concept, Eq.~(\ref{eq:zeroQ}) and Eq.~(\ref{eq:qW}), is applicable to general closed and open dynamical processes \cite{Jarzynski04,CTH09}. Thus they appear to refer to \emph{different types of work}, a situation that may be compared to different entanglement measures each of importance for a different quantum communication and computational tasks. For example, it has been suggested that the resource theory work is a suitable measure to quantify \emph{resonance processes} \cite{GA15}. A second issue is the link between the work definition in the ensemble sense, see (\ref{eq:Q+W}), and the single shot work. Beyond taking mathematical limits of Renyi entropies these definitions have to be understood within the context of each other operationally, i.e. how can one measure each of these quantities and in what sense do these quantities converge experimentally? Indeed, while \ja{traditional classical} thermodynamics is a manifestly practical theory, made for steam engines and fridges, quantum thermodynamics has only made a small number of experimentally checkable predictions of new thermodynamic effects yet. 

Another point of discussion is the kinematic versus dynamical approach to thermalisation and equilibration. The kinematic approach arises from typicality discussions in quantum information where states in Hilbert space are discussed from the standpoint of a property of their marginals. In the case of thermalisation, this property is the closeness to the canonical Gibbs state. A Hamiltonian and its eigenstates never appear in the kinematic description. The situation is quite different in the case of the dynamical approach to thermalisation implied by the eigenstate thermalisation hypothesis (ETH), where the eigenstates of the Hamiltonian are of crucial importance. A connection between these two techniques to study thermalisation is thus desirable. This difference in perspectives between quantum information theoretic and kinematic approaches also needs to be reconciled in the study of pre-thermalisation. Exceptional states such as ``rare states'' need to be fully understood from a kinematic standpoint, see \cite{biroli2010effect} for a discussion on rare states and thermalisation.

\add{
In the context of quantum thermal machines, the interplay between statistics and engine performance needs more study \cite{PhysRevE.90.062121,zheng2014work}. This is crucial since typical work fluids of quantum thermal machines constitute several particles. Though some authors have considered non-thermal baths in contact with a work fluid, almost all such work involves unitary transformations on thermal states. There is a need for the study of thermal machines wherein at least one of the baths is non-thermal in a way that is different from unitary transformations on thermal baths. The role of entanglement and correlations would clearly become very important in such a regime. Finally, we note that various experimental implementations of quantum engines are expected in the near future \cite{rossnagel16}.}

\jan{To close, quantum thermodynamics is a rapidly evolving research field that promises to change our understanding of the foundations of physics while enabling the discovery of novel thermodynamic techniques and applications at the nanoscale. This overview provided an introduction to a number of current trends and perspectives in quantum thermodynamics and concluded with three particular discussions where there is still disagreement and flux. Resolving these and other riddles will no doubt deepen our understanding of the interplay between quantum mechanics and thermodynamics.}


\section*{Acknowledgements} 
We are grateful to A. Acin, A. Auffeves, G. De Chiara, V. Dunjko, J. Eisert, R. Fazio, A. Fisher, I. Ford, P. Gonzalo Manzano, M. Horodecki, K. Hovhannisyan, M. Huber, K. Jacobs, D. Jennings,  R. Kosloff, M.A. Martin-­Delgado, M. Olshanii, M. Paternostro, M. Perarnau-Llobet, M. Plenio, D. Poletti, S. Salek, P. Skrzypczyk, R. Uzdin and M. Wilde for their input on the overview. The Centre for Quantum Technologies is a Research Centre of Excellence funded by the Ministry of Education and the National Research Foundation of Singapore. JA acknowledges support by the Royal Society and EPSRC (EP/M009165/1). This work was supported by the European COST network MP1209 ``Thermodynamics in the quantum regime''.


\end{document}